\documentclass[%
 reprint,
superscriptaddress,
 amsmath,amssymb,
 aps,
 prl,
]{revtex4-2}
\usepackage{hyperref}
\usepackage{epstopdf,epsfig}  
\usepackage{physics}
\usepackage{soul}
\usepackage{color}
\usepackage{graphicx}
\usepackage{dcolumn}
\usepackage{bm}
\usepackage{algorithm}
\usepackage{algpseudocode}
\newcommand{\sectionnotoc}[1]{
\begin{center}
    { \normalsize\textsc{\textbf{#1}}}
\end{center}
}

\begin{document}

\title{Modeling collective behaviors from optic flow and retinal cues}

\author{Diego Castro}

\affiliation{
Aix Marseille Université, CNRS,  ISM, Marseille, 13288, France
}
 \affiliation{
Aix Marseille Université, CNRS, Centrale Med, IRPHE, Marseille, 13013, France
}
\author{Franck Ruffier}%
\email{franck.ruffier@cnrs.fr}
\affiliation{
Aix Marseille Université, CNRS,  ISM, Marseille, 13288, France
}
\author{Christophe Eloy}
 \affiliation{
Aix Marseille Université, CNRS, Centrale Med, IRPHE, Marseille, 13013, France
}

\date{\today}

\begin{abstract}
Animal collective behavior is often modeled with self-propelled particles, assuming each individual has ``omniscient'' knowledge of its neighbors. Yet, neighbors may be hidden from view and we do not know the effect of this information loss. To address this question, we propose a visual model of collective behavior where each particle moves according to bio-plausible visual cues, in particular the optic flow. This visual model successfully reproduces three classical collective behaviors: swarming, schooling, and milling. This model offers a potential solution for controlling artificial swarms visually. 
\end{abstract}

\maketitle
\sectionnotoc{INTRODUCTION}

Collective animal behavior is a widespread phenomenon in nature, ranging from the mesmerizing movements of starling murmurations to the coordinated motion of cattle herds \cite{liebchen2017collective,caussin2014emergent,barberis2016large,silverberg2013collective}. These collective behaviors are commonly modeled with self-propelled particles: individuals with an intrinsic speed orient themselves based on a set of rules, alignment, attraction, and avoidance \cite{vicsek1995novel,reynolds1987flocks,ginelli2010relevance,albano1996self}. These ``3A'' rules have been successful in reproducing different collective behaviors, such as swarming (no orientational order), schooling (high orientational order), and milling (the group coordinately swirls in a circular pattern)  \cite{katz2011inferring,mishra2010fluctuations,solon2015pattern,vicsek2012collective}. Furthermore, numerous studies used these rules to replicate some of these phases on artificial robotic swarms \cite{viragh2014flocking, wang2019collective,shen2022multi}.

In self-propelled particle (SPP) models , the rules of attraction, alignment, and avoidance are typically applied with the simplifying assumption that each individual possess idealized senses, is "omniscient" and gauges perfectly the position, distance, orientation, and velocity of its neighbors \cite{bode2010perceived,mateo2019optimal}. However, this assumption may not hold in practice, especially when some neighboring individuals are hidden from view \cite{soria2019influence,davidson2021collective,strandburg2013visual,lemasson2013motion,ballerini2008interaction}.

The causal link between visual cues and collective behavior has been shown through several vision-based biological models \cite{lemasson2009collective, collignon2016stochastic,qi2022emergence,Aoki1984Dynamics}. As a result, different ways of incorporating vision in SPP models have been suggested. The most widespread approach consists of using vision to filter information \cite{krongauz2024vision, ariel2014individual}. One of these models suggested that flocks of starlings adjust their density to reach a state of ``marginal opacity'' \cite{pearce2014role}. However, this density adjustment does not seem to be widespread, as certain animals such as fish can form opaque schools \cite{ballerini2008empirical,ginelli2010relevance}.

In a recent study, Bastien and Romanczuk proposed a model of collective behavior based purely on vision capable of reproducing most of the collective behaviors \cite{bastien2020model}.  Their model simulates each individual's response to a projection of the visual field, rather than relying on omniscient information. However, the portrayed milling phase is uncoordinated, meaning that the particles turn in both directions in the same swirl. And, while they claim to use the simplest possible equations of movement that satisfy fundamental symmetries, their model involves six parameters that are hard to relate to the classical 3A rules. In addition, four terms of their model involve spatial and temporal derivatives that would use important computing resources. Instead, we propose to use bio-plausible visual cues that can be measured directly by a visual sensor, i.e. the optic flow and the retinal position.

Optic flow refers to the apparent angular velocity of objects in the visual field due to the relative motion between the observer and its surroundings. Numerous animal species perceive and use optic flow for a variety of tasks. Insects use it to navigate in crowded settings \cite{portelli2010honeybees}, evade ground obstacles, and control their landing \cite{franceschini2007bio}.  Fish use it for navigation \cite{bonnen2023motion}, and birds use it during takeoff \cite{serres2019optic}. Optic flow is ubiquitous in nature. It involves specialized neurons, well-identified in invertebrates, which have inspired bio-inspired sensors dedicated to optic flow \cite{ruffier2005optic, expert2015flying, floreano2013miniature,  schilling2022scalability}. These visual sensors can provide panoramic optic flow sensing \cite{floreano2013miniature, expert2015flying} and allows for direct measurement and on-board panoramic use of this visual cue \cite{bergantin2021oscillations, de2022accommodating}.

In this article, we propose a non-stereoscopic vision-based model of collective behavior, inspired by animal vision. This model is intended to be implementable on autonomous robots equipped with visual sensors, i.e. the robots do not need to communicate with a central command or/and to know their neighbors' relative coordinates, their relative positions between each other or to be georeferenced. Our model aims at bridging the gap between traditional SPP models that rely on omniscient information and biomimetic visual approaches.
\sectionnotoc{MODEL}
\begin{figure}
  \centering
  \includegraphics[width=\columnwidth]{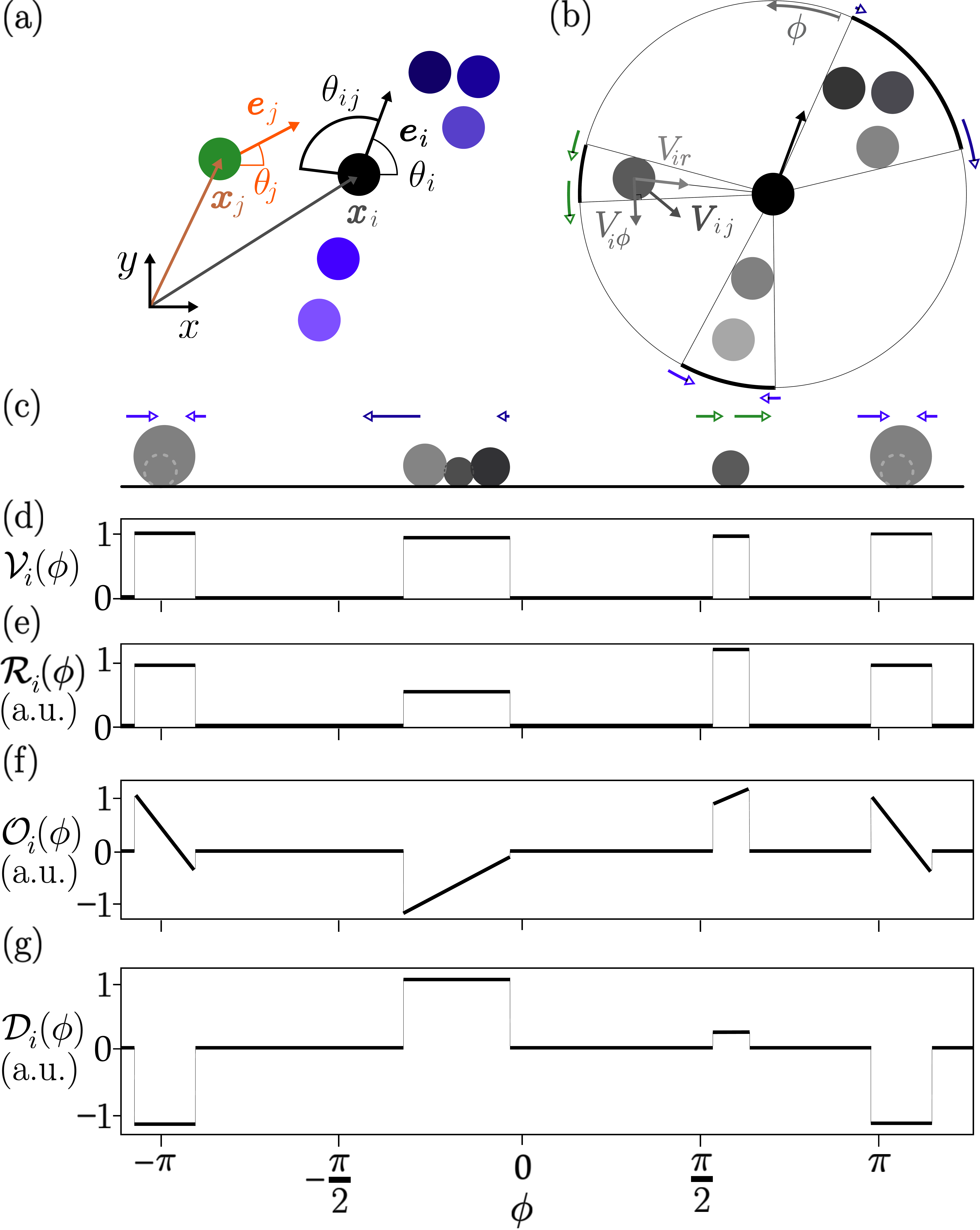}
  \caption{
Notations used and principle of the visual cues. 
  (a) The particle $i$ is located in $\boldsymbol{x}_i$ and $\theta_i$ is its heading. $\theta_{ij}$ is the retinal position of particle j for the $i$-th particle.
  (b) The angle $\phi$ is the line of sight with respect to the particle's heading. The vector $\boldsymbol{V}_{ij} = U(\boldsymbol{e}_j-\boldsymbol{e}_i)$ is the relative velocity of particle $j$ perceived by $i$. It can be decomposed into radial and azimuthal components, $V_{ir}$ and $V_{i\phi}$.
  (c) Point of view of the particle $i$, the arrows represent the optic flow generated by the relative angular velocity of each particle.
  (d) The visual field $\mathcal{V}_i(\varphi)$ is a binary representation of (c). It is formed of a set of ``shades''.  
  (e) The function $\mathcal{R}_i(\phi)$ represents the estimated distance associated to shades, assuming they are associated with a single individual of radius $a$.
  (f) Optic flow $\mathcal{O}_i(\phi)$ associated with the apparent angular velocity of the shade edges.
  (g) Optic flow divergence $\mathcal{D}_i(\phi)$.
  }
  \label{fig:OcclusionMapCons}
\end{figure}
We consider a system of $N$ self-propelled particles in two dimensions. Each particle is a circular object with radius $a$, moving with a constant speed $U$. The position of the $i$th particle is noted $\boldsymbol{x}_i$ and its direction  $\boldsymbol{e}_i=[\cos \theta_i, \sin \theta_i]$, with $\theta_i$ its heading [Fig.~\ref{fig:OcclusionMapCons}(a)]. 

We model the interactions between particles using changes in their angular velocity associated with attraction, alignment, and noise. Specifically, the equations of motion can be written as follows
\begin{subequations}
\label{eq:GeneralForm}
\begin{eqnarray}
\dot{\boldsymbol{x}}_i & = & U \boldsymbol{e}_i, \\ 
\dot \theta_i & = & k_\odot \, \omega_\odot + k_\parallel \, \omega_\parallel + k_\eta \, \eta,
\end{eqnarray}
\end{subequations}
where dots denote temporal derivatives and $\eta(t)$ is a standard Wiener process representing rotational noise. The functions $\omega_\odot$ and $\omega_\parallel$ are $O(1)$ functions representing attraction and alignment. The parameters $k_\odot$, $k_\parallel$,  $k_\eta$ control the strength of attraction, alignment, and noise. For simplicity, we did not include an avoidance rule, as it is not a required rule to reproduce collective behaviors \cite{caussin2014emergent,morin2015collective,vicsek2012collective}.

We begin by introducing an omniscient model that will serve as a reference. This model is inspired by a data-driven fish model \cite{calovi2014swarming,filella2018model}, with the difference that each particle interacts with all the others. The attraction and alignment terms are given by  
\begin{subequations}\label{eq:omni}
\begin{eqnarray}
\omega_\odot^\mathrm{omni.} & = &
    \left\langle \sum_{j=1}^{N}
        \|\boldsymbol{x}_j-\boldsymbol{x}_i\|\sin{ (\theta_{ij})}b_\epsilon(\theta_{ij})\right\rangle,
        \label{eq:omniattraction} \\
\omega_\parallel^\mathrm{omni.} & = &
	\left\langle \sum_{j=1}^N
        \frac{\boldsymbol{e}_i \times \boldsymbol{e}_j}{\|\boldsymbol{x}_j-\boldsymbol{x}_i\|^2}  b_\epsilon(\theta_{ij})\,
    .\right\rangle,
\label{eq:omnialignment}        
\end{eqnarray}
\end{subequations}
where $b_\epsilon(\phi) = 1 + \epsilon \cos {\phi}$ models the blind angle (See Also Fig. S1, \cite{SuppMateria}). When $\epsilon=0$, $b_\epsilon$ is isotropic; when $\epsilon=1$, the particle cannot see behind itself \cite{SuppMateria}. The brackets denote a normalization defined as
\begin{equation}\label{eq:normalization}
\left\langle \sum_j
     f(j) \sin \theta_j \right\rangle =      {\sum_j f(j)\sin \theta_j }\bigg/{\sum_j |f(j)|}
. 
\end{equation}

To model visual perception, we assume that each particle senses a visual field $\mathcal{V}_i(\phi)$, where $\phi$ represents the angle between the particle’s heading and the line of sight [Fig.~\ref{fig:OcclusionMapCons}(c)]. The function output is binary, indicating the presence or absence of a shade in the visual field [Fig.~\ref{fig:OcclusionMapCons}(b)-(d)]. 

Using the information from the visual field $\mathcal{V}_i(\phi)$, we can derive the function $\mathcal{R}_i(\phi)=a/\sin(\Delta\phi/2)$, where $\Delta\phi$ represents the angle of view angle of shades [Fig.~\ref{fig:OcclusionMapCons}(e)]. With this definition, when a shade is associated with a single particle, $\mathcal{R}_i(\phi)$ represents its distance from the viewer. 

Temporal changes in the features of the visual field can be used to calculate the optic flow $\mathcal{O}_i(\phi)$. A simplified optic flow is estimated by assuming that each shade has a pattern that moves and deforms with it. It results that $\mathcal{O}_i(\phi)$ is a linearly interpolated function of the angular velocity between two features of the shade, its rising and falling edges [Fig.~\ref{fig:OcclusionMapCons}(f)]. 
This method to compute the optic flow computation corresponds to a cross-correlation of visual field features \cite{ullman1981analysis}, which are known to occur in animal eyes  \cite{franceschini1989directionally,eichner2011internal}.
Similarly, we can compute the optic flow divergence $\mathcal{D}_i(\phi)$ by derivating the optic flow $\mathcal{O}_i(\phi)$ [Fig.~\ref{fig:OcclusionMapCons}(f)]. Due to the piece-wise linear nature of $\mathcal{O}_i(\phi)$, $\mathcal{D}_i(\phi)$ is a piece-wise constant function.

The optic functions $\mathcal{V}$, $\mathcal{R}$, $\mathcal{O}$, and $\mathcal{D}$ are inspired by animal vision. These functions can easily be computed by a man-made vision system. We will now use these functions to the attraction and alignment terms of a vision model.

In this visual model, the attraction and alignment terms are given by 
\begin{subequations}
\begin{eqnarray}
\omega_{\odot}^\mathrm{visu.} & =& \left\langle 
    \int_{-\pi}^\pi 
        \mathcal{R}_i^2(\phi)  b_\epsilon(\phi) \sin \phi\,
    d\phi
\right\rangle,
\label{eq:attraction} \\
\omega_\parallel^\mathrm{visu.} & =&  \left\langle 
    \int_{-\pi}^\pi 
        \frac{\boldsymbol{e}_i \times \boldsymbol{V}_{ij}}{U\mathcal{R}_i(\phi)} b_\epsilon(\phi)\,
    d\phi.\right\rangle,\label{eq:alignment}
\end{eqnarray}
\end{subequations}
where the brackets denote the normalization given by Eq.~\eqref{eq:normalization} with the sum replaced by the integral. The attraction and alignment terms of the visual model are constructed to be similar to those of the omniscient model given in Eqs.~(\ref{eq:omni}a)-(\ref{eq:omni}b). The difference in the exponent of $\mathcal{R}$ comes from the additional $\Delta \phi \sim 1/\mathcal{R}$ arising from the integration.  

In the alignment term, $\omega_\parallel^\mathrm{visu.}$, the cross-product $\boldsymbol{e}_i \times \boldsymbol{e}_j$ is evaluated from the visual information. This is done by using the optic flow $\mathcal{O}_i(\phi)$ and its divergence $\mathcal{D}_i(\phi)$, which are related to the velocity of particle $j$ with respect to particle $i$ [Fig.~\ref{fig:OcclusionMapCons}(b)]. Specifically, the radial component is given by $V_{ir}=-\mathcal{R}_i(\phi)$ $\mathcal{D}_i(\phi)$, and the azimuthal component by $V_{i\phi}=\mathcal{R}_i(\phi)$ $\mathcal{O}_i(\phi)$. We can use these components to calculate the vector $\boldsymbol{V}_{ij}$ in polar coordinates $(r, \phi)$ as $\boldsymbol{V}_{ij}=(-\mathcal{D}_{i}, \mathcal{O}_{i})\mathcal{R}_{i}/U$, while $\boldsymbol{e}_i$ can be expressed as $(\cos \phi, -\sin \phi)$. It results that
\begin{equation}\label{eq:ei_ej}
\frac{\boldsymbol{e}_i \times \boldsymbol{V}_{ij}}{U\mathcal{R}_i(\phi)} = \frac{-\mathcal{D}_i(\phi) \sin\phi + \mathcal{O}_i(\phi) \cos\phi}{U}.
\end{equation}

The alignment term is thus the sum of two terms: one proportional to the derotated optic flow $\mathcal{O}$ and sensitive to the azimuthal velocity of neighbors, and one proportional to the optic flow divergence $\mathcal{D}$ and sensitive to the radial velocity. 
When computing $\boldsymbol{e}_i \times \boldsymbol{V}_{ij}$ with Eq.~\eqref{eq:ei_ej}, we remove the particle rotation with angular velocity $\dot\theta_i$ from the optic flow $\mathcal{O}_i(\phi)$.

The equations of motion presented in Eq.~\eqref{eq:GeneralForm}, along with the attraction and alignment terms derived in Eqs.~\eqref{eq:attraction} and \eqref{eq:alignment}, provide a model of collective behavior based on realistic visual cues. To make the problem dimensionless, we chose $a = 1$ and $U = 1$. With this approach, four dimensionless parameters remain, the strengths of noise attraction and alignment, $k_\eta$, $k_\odot$, $k_\parallel$, and the blind angle parameter $\epsilon$. 
\sectionnotoc{NUMERICAL SIMULATIONS}

\begin{figure}[t]
  \centering
  \includegraphics[width=1\columnwidth]{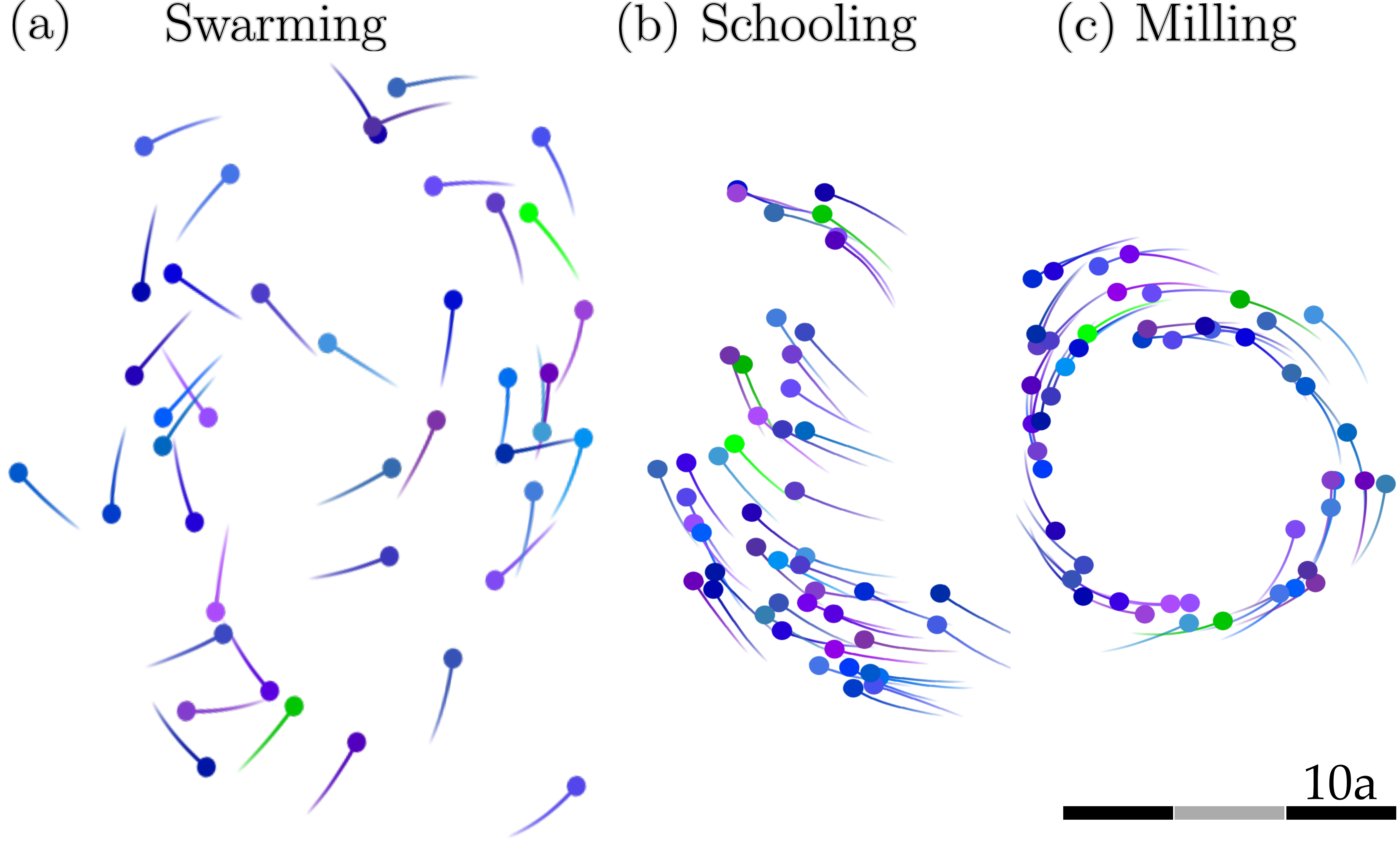}
  \caption{Illustration of the phases observed for $N=50$ individuals. The scale bar measures $30a$ in total ($10a$ for each section). The parameters $\epsilon=1$ and $k_\eta=0.01$ are fixed. We observe three phases when varying the other two parameters:
 (a) swarming ($k_\odot=0.1$, $k_\parallel=0$);
 (b) schooling ($k_\odot=0.06$, $k_\parallel=0.2$); and
 (c) milling ($k_\odot=0.1$, $k_\parallel=0.04$).}
\label{fig:phases}
\end{figure}

To explore the effect of these parameters on collective behaviors, we performed numerical simulations with $N=50$ particles. 
Initially, the particles are randomly placed in a square of side $aN$ with random headings \cite{SuppMateria}\footnote{The open source code will be accessible upon acceptance of this article}. 
The dynamical system described by Eqs.~(\ref{eq:GeneralForm}a) and (\ref{eq:GeneralForm}b) is solved numerically using a discrete implementation of Eqs.~(\ref{eq:attraction}) and (\ref{eq:attraction}) (See also Figs. S2-S3, \cite{SuppMateria}).
We examined the effect of the time step $\delta t$ by conducting simulations with  $\delta t =$ 0.001, 0.01, and 0.1 (Figs.~S17--S22 and Video S1-S6,  \cite{SuppMateria}). However, no significant differences were observed on the collective behavior, and a time step of $\delta t$ = 0.1 was selected for the remaining simulations to ensure computational efficiency. 

We first set $k_\eta=0.01$ and $\epsilon=1$ (maximum blind angle) and explore the effects of the two remaining parameters $k_\odot$ and $k_\parallel$ in the visual model. Our simulations show three distinct dynamical phases (Fig.~\ref{fig:phases} for the visual model, Fig.~S10 for the omniscient model and Figs.~S23--S28 snapshots for the evolution of the stable phases seen on Videos~S7-S9 and S10-S12, \cite{SuppMateria}). If the alignment is zero, a disordered swarming phase is observed, where individuals form a group without a preferred direction [Fig.~\ref{fig:phases}(a)]. When the alignment strength increases, particles begin to align in the same direction, resulting in the schooling phase [Fig.~\ref{fig:phases}(b)]. If the ratio between the alignment and the attraction strengths is around $0.4$, the group exhibits a milling phase [Fig.~\ref{fig:phases}(c)], creating a vortex. These three phases (swarming, schooling, and milling) have regularly been observed in (omniscient) self-propelled-particle models \cite{calovi2014swarming,filella2018model,charlesworth2019intrinsically,lu2022improved,shang2014influence}  

To quantitatively distinguish between the different dynamical phases, we introduce three global order metrics: polarization $P$, milling $M$, and opacity $O$ \cite{calovi2014swarming,filella2018model}. These metrics are defined as follows,
\begin{subequations}
\begin{eqnarray} \label{eq:Metrics}
P &=&\| \overline{\boldsymbol{e}_i} \|, \\ 
M &=& \| \overline{\boldsymbol{y}_i \times \boldsymbol{e}_i} \|, \\
O &=& \frac{1}{2\pi}\overline{\int^{\pi}_{-\pi}\mathcal{V}_i(\phi)d\phi},
\end{eqnarray}
\end{subequations}
where the overbar represents an average over all individuals and the unit vector $\boldsymbol{y}_i= (\boldsymbol{x}_i- \overline{ \boldsymbol{x}_i})/\|\boldsymbol{x}_i- \overline{ \boldsymbol{x}_i}\|$ points towards particle $i$ from the center of mass. All three metrics range in the interval $[0, 1]$. The polarization $P$ measures the alignment: $P=0$ corresponds to particles pointing in all directions, $P=1$ corresponds to a perfectly aligned school. The milling $M$ represents the normalized angular momentum: straight-line formation gives $M=0$ and perfect milling gives $M=1$. The opacity $O$ measures the ``occupancy'' of the visual fields: $O=0$ when there is no object in the visual field, and $O=1$ when the entire visual field is obscured.

\begin{figure*}
  \includegraphics[width=2\columnwidth]{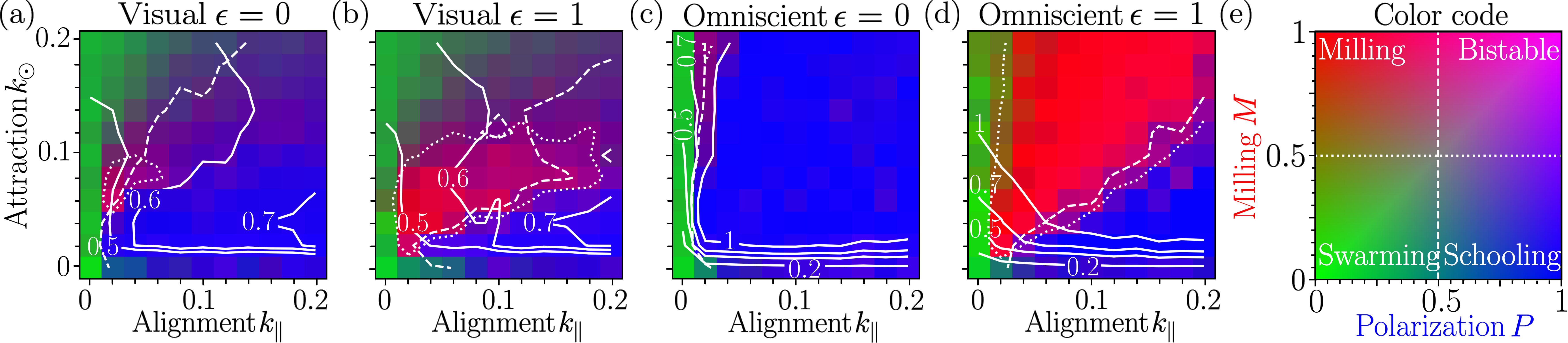}
  \caption{Phase diagrams for $N=50$ individuals and noise strength $k_\eta=0.01$. These phase diagrams compare the visual model (a), (b) and the omniscient model (c), (d) for $\epsilon=0$ (a), (c) and $\epsilon=1$ (b), (d). The colors represent different values of $P$ and $M$ as shown in (e). The contours show the values of opacity $O$ (white solid), the line $M=0.5$ (white dotted), and $P=0.5$ (white dashed).}
  \label{fig:mapping}
\end{figure*}

We now compare the visual and omniscient models by setting the value of the noise to $k_\eta=0.01$ and exploring the parameter space $(k_\odot, k_\parallel) \in [0,0.2]\times[0,0.2]$ for two values of the blind angle parameter ($\epsilon=0$ or 1).  
For each parameter set, we ran 10 simulations over long durations ($\Delta t = 5000$).  
The mean values of $P$, $M$, and $O$ were determined by averaging over the last 1000 time units to ensure that the transient has no influence. The outcomes of these simulations are synthesized in phase diagrams [Fig.~\ref{fig:mapping},  and expanded on each metric independently on Figs.~S4,~S6--S9,~S11,~	S13-S16. \cite{SuppMateria}].
 \sectionnotoc{DISCUSSION}
Somewhat arbitrarily, we chose to identify the collective phases from the values of the polarization $P$ and the milling $M$ parameters: schooling when $P>0.5$ and $M<0.5$; milling when $P<0.5$ and $M>0.5$; swarming when $P<0.5$ and $M<0.5$; and bistable when $P>0.5$ and $M>0.5$ (we will come back to this particular phase below). 

Our vision model qualitatively reproduces the phases observed in the omniscient model [Fig.~\ref{fig:mapping}]. Specifically, we observe the three phases in the vision model: schooling when $k_\odot\lesssim 0.5 k_\parallel$; swarming when $k_\parallel \lesssim 0.3 k_\odot$; and milling or bistability otherwise. When the blind angle parameter increases, it tends to stabilize the milling phase both in the visual model and the omniscient model, as observed in the literature \cite{calovi2014swarming}.

The difference between the two models increases at high values of the alignment and attraction strengths when the opacity $O$ is maximum. 
In the visual model, the opacity does not exceed $0.7$, regardless of the values of $\epsilon$ and $k_\eta$. In the omniscient model, however, the opacity exceeds $0.8$ and even reaches one [Fig.~\ref{fig:mapping}(c) and (d)]. This is because the radius $a$ does not play any role in the omniscient model and the characteristic length is set by $k_\eta^2$. In both models, as expected, larger noise strength causes a decrease in average opacity (Figs. S4 and S11)\cite{SuppMateria}.  

\begin{figure}[t]
  \centering
  \includegraphics[width=0.95\columnwidth]{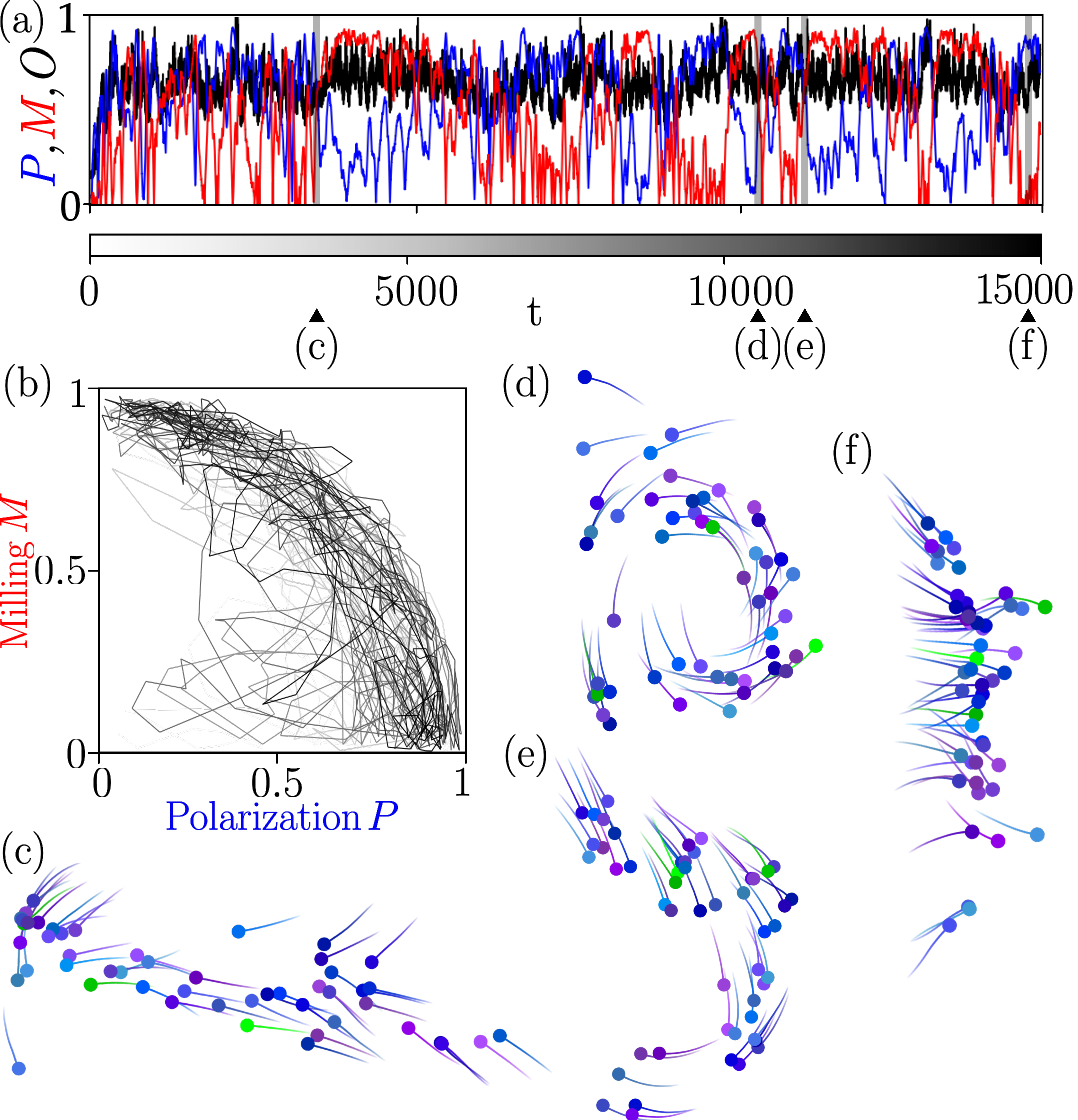}
  \caption{
  Example of the bistability observed in the visual model (parameter values: $\epsilon=1$, $k_\odot=0.1, k_\parallel=0.2$, and  $k_\eta=0.01$). 
(a) Time series of the three metrics $P$, $M$, and $O$. 
(b) Trajectories of the collective behavior in the $(M,P)$-plane.
Illustration of the phases:
(c) schooling-milling transition; (d) milling; (e) milling-schooling transition; and (f) schooling.
}
  \label{fig:bistability}
\end{figure}

Now, let us examine the bistable phase. 
Figures \ref{fig:bistability} and S29 show this phase in the visual model for $\epsilon=1$, $k_\odot=0.1, k_\parallel=0.2$, and  $k_\eta=0.01$. After a transient, the group forms a milling phase until $t\approx 6000$, but eventually it transitions to a schooling phase for $6000\lesssim t\lesssim 7500$ before returning to a schooling phase again, and so on.   
These transitions show that the system exhibits a noise-induced intermittency between two stable states: milling and schooling.
The schooling phase far from the transition resembles a front of parallel individuals [Fig.~\ref{fig:bistability}(f)]. Just before the transition to milling, some individuals move ahead, [Fig.~\ref{fig:bistability}(c)]. Reciprocally, the milling phase just before the transition to schooling opens up, generating a C shape [Fig.~\ref{fig:bistability}(e)].
These transitions are fairly stereotyped as they tend to follow the same path in the $(M,P)$ plane [Fig.~\ref{fig:bistability}(b)].
The existence of these intermittent transitions mediated by noise is likely due to a second-order transition between the milling and schooling phases as seen on the metric distribution along the boundaries of both phases [Figs.~S6-S9 and S13-S16]. This multi-stability has been present on similar 3A models \cite{calovi2014swarming}.

We changed the group size from $N=5$ to $300$ (Figs.~S5 and S12, \cite{SuppMateria}). Although the group size does not seem to impact qualitatively the phase diagram, small groups tend to favor schooling, whereas large groups tend to favor swarming. 

\sectionnotoc{CONCLUSION}

In conclusion, we proposed a model based on biologically plausible visual cues. This model successfully reproduces the three classical phases of animal collective behavior: swarming, schooling, and milling. These findings show that visual cues provide enough information to enable collective behavior. 

Furthermore, our findings imply potential practical uses for synchronizing groups of artificial drones, which may be governed by analogous visual stimuli. In future studies, we aim to investigate these opportunities more thoroughly and enhance our model for a more accurate depiction of animal collective behavior in a three-dimensional space.
\let\oldaddcontentsline\addcontentsline
\renewcommand{\addcontentsline}[3]{}

\let\addcontentsline\oldaddcontentsline
\clearpage

\renewcommand{\figurename}{Fig. S}

\newcounter{Cequ}
\newcounter{Caux}

\newenvironment{CEquation}
  {\stepcounter{Cequ}%
    \addtocounter{equation}{-1}%
    \renewcommand\theequation{S\arabic{Cequ}}\equation}
  {\endequation}
\newenvironment{CAlign}
  {\setcounter{Caux}{\theequation}
    \setcounter{equation}{\theCequ}%
    \renewcommand\theequation{S\arabic{equation}}
    \align}
  {\endalign\setcounter{Cequ}{\value{equation}}\setcounter{equation}{\theCaux}}

\newcounter{Nsubsection}
\newcounter{Nsubsubsection}
\counterwithin{Nsubsection}{section}
\counterwithin{Nsubsubsection}{Nsubsection}
\newcommand{\sectionmy}[1]{
\stepcounter{section}
\section{S\arabic{section} #1}
}
\newcommand{\subsectionmy}[1]{
\stepcounter{Nsubsection}
\subsection{S\arabic{section}.\arabic{Nsubsection} #1}
}
\newcommand{\subsubsectionmy}[1]{
\stepcounter{Nsubsubsection}
\subsection{S\arabic{section}.\arabic{Nsubsection}.\arabic{Nsubsubsection} #1}
}

\newcommand{\dtcaption}[2]{
Snapshot of the time step stability at $\delta$t=#1 for the #2 model. Reproduction of a milling phase for a flock size of 50 individuals, $k_\odot=$0.06,$k_\parallel=$0.06 and $k_\eta=$0.01. a) Evolution of the flock metrics, b-j different snapshots.
}
\newcommand{\variancecaption}[4]{
Phase diagram for the #1 model of a flock size of #2 individuals, $k_\eta=$#3 and #4 blind angle. a) Integrated metrics, b) Separated metrics and c) Variance between the 10 experiments final metrics.}
\newcommand{\phasecaption}[7]{
#1 phase reproduction for a flock size of #2 individuals for the #3 model.  $k_\eta=$#4 $k_\eta=$#5 $k_\eta=$#6 and #7 blind angle
}

\sectionnotoc{Supplemental Material for \\ ``Modeling collective behaviors from optic flow and retinal cues" \\ D. Castro, F. Ruﬀier, and C. Eloy}
\tableofcontents
\sectionmy{Mathematical development}
\subsectionmy{Visual and Omniscient attraction equivalence proof}

We defined the attraction rule for the omniscient and visual models to be equivalent one to the other. But changing the discrete sum of the omniscient model into an integral needed for the visual model, requires to multiply the integrand by  $\mathcal{R}_i(\phi)$. 

Here we present a step by step calculation of this integral to show the equivalence with the discrete sum. We start by the integral of Eq. (4a),
\begin{CEquation}
    \omega^\mathrm{visu.}_\odot =  
    \left\langle\int_{-\pi}^\pi 
        \mathcal{R}^2_i(\phi)b_\epsilon(\phi) \sin \phi\,
    d\phi,\right\rangle
    \label{eq:Attraction}
\end{CEquation}
where $b_\epsilon(\phi) = 1 + \epsilon \cos \phi$. 

$\mathcal{V}_i(\phi)$ is a piece-wise constant function that can be expressed as a set unit boxcar functions (a shade). Each, can be described as a rising $\varphi_{\Uparrow k}$ and its corresponding falling edge $\varphi_{\Downarrow k}$. Furthermore, from these edges, we can compute the shade's mid-point $\varphi_k = (\varphi_{\Uparrow k}+\varphi_{\Downarrow k})/2$, and the shade's  half-width $\Delta \varphi_k =( \varphi_{\Downarrow k} - \varphi_{\Uparrow k})/2$. $\mathcal{R}_i(\phi)$ is $\mathcal{V}_i(\phi)$ weighted by $\frac{a}{\sin\Delta\varphi_k}$. This means that  $\mathcal{R}_i(\phi)$ for a binary $\mathcal{V}_i(\phi)$, is independent of $\phi$ and is only be dependent on the $k$-shade's half-width ($\mathcal{R}_k(\Delta \varphi_k)$). The subscript changes from $i$ to $k$, as $i$ indicate is a value referred to continuous information, k to discrete information of a single shade. A shade k id different from an individual $j$, as in a shade can be occlusion or aggregation of any $j$-particles. Then, we can rewrite Eq.~\eqref{eq:Attraction} as:

\begin{CEquation}
    \label{eq:attstep1}
    \omega^\mathrm{visu.}_\odot=\left\langle\sum_k \mathcal{R}^2_k(\Delta\varphi_k)\int^{\varphi_{\Downarrow k}}_{\varphi_{\Uparrow k}}  b_\epsilon(\phi) \sin \phi\,
    d\phi.\right\rangle
\end{CEquation}
\begin{CEquation}\label{attractionm}
\begin{matrix}
    \omega^\mathrm{visu.}_\odot=\langle\sum_k 2 \mathcal{R}^2_k(\Delta\varphi_k) \sin(\Delta\varphi_k) \sin(\varphi_k) \\ ~ ~ \left(
        1 + \epsilon\cos\varphi_k\cos\Delta\varphi_k
    \right) \rangle
\end{matrix}
\end{CEquation}
And here $\mathcal{R}_k(\Delta\varphi_k)=\frac{a}{\sin\Delta\varphi_k }$, Resulting in:

\begin{CEquation}
    \label{eq:attstep1}
    \omega^\mathrm{visu.}_\odot=\left\langle\sum_k 2\frac{a^2\sin(\Delta\varphi_k)}{\sin^2(\Delta\varphi_k)}  \sin(\varphi_k) \left(
        1 + \epsilon\cos\varphi_k\cos\Delta\varphi_k
    \right)\right\rangle
\end{CEquation}
\begin{CEquation}\label{eq:S5}
    \omega^\mathrm{visu.}_\odot=\left\langle\sum_k 2a\mathcal{R}_i(\Delta\varphi_k) \sin(\varphi_k) \left(
        1 + \epsilon\cos\varphi_k\cos\Delta\varphi_k
    \right)\right\rangle
\end{CEquation}
For the case where there is no occlusion nor aggregation $k=j$ and taking into account the physical constrains of the particles: $\cos\Delta\varphi_k \approx 1$, $\mathcal{R}_i(\Delta\varphi_k)=||\boldsymbol{x}_j-\boldsymbol{x}_i||$ and $\varphi_k=\theta_{ij}$, then Eq.~\eqref{eq:S5} becomes 
\begin{CEquation}\label{eq:S6}
    \omega^\mathrm{visu.}_\odot\approx\left\langle\sum^N_{j\ne i} 2a||\boldsymbol{x}_j-\boldsymbol{x}_i|| \sin(\theta_{ij}) \left(
        b_\epsilon(\theta_{ij})
    \right)\right\rangle
\end{CEquation}
Finally the $2a$ factor is nullified by the $\langle .\rangle$ normalization. So it becomes identical to the attraction term in Eq. (2a), proving the equivalence between the two models.

\subsectionmy{Normalization for non-trigonometric equation}

We presented a normalization to guarantee the O(1) for all the terms of the models. Here we proof that the the non-trigonometric equation presented that are normalized by this equation, does in fact maintains the same form as shown on Eq. (3). We start by taking the alignment term of the omniscient model,

\begin{CEquation}\label{eq:S7}
\omega_\parallel^\mathrm{omni.}  = 
	\left\langle \sum_{j\ne i}^N
        \frac{\boldsymbol{e}_i \times \boldsymbol{e}_j}{\|\boldsymbol{x}_j-\boldsymbol{x}_i\|^2}  b_\epsilon(\theta_{ij})\,
    .\right\rangle,
\end{CEquation}
this cross product can be expressed as:

\begin{CEquation}
        \boldsymbol{e}_i\cross \boldsymbol{e}_j = U\|\boldsymbol{e}_i\| \|\boldsymbol{e}_j\| \sin (\theta_i-\theta_j)
    \label{eq:velocityVectorscross}
\end{CEquation}

Rewriting equation (2b) to better isolate the terms of the normalization described on equation (3)
\begin{CEquation}\label{eq:S9}
\omega_\parallel^\mathrm{omni.}  = 
    \left\langle \sum_{j=1}^N
        \frac{U\|\boldsymbol{e}_i\| \|\boldsymbol{e}_j\| }{\|\boldsymbol{x}_j-\boldsymbol{x}_i\|^2}  b_\epsilon(\theta_{ij}) \sin (\theta_i-\theta_j) 
    ,\right\rangle,
\end{CEquation}
where, $\boldsymbol{e}_i$ and $\boldsymbol{e}_j$ are a unit vectors. Here $\sin(\theta_j)=\sin (\theta_i-\theta_j) )$ and $f(j)=\frac{U\|\boldsymbol{e}_i\| \|\boldsymbol{e}_j\| }{\|\boldsymbol{x}_j-\boldsymbol{x}_i\|^2}  b_\epsilon(\theta_{ij})$. Which maintains the form of the normalization presented on Eq. (3).

\subsectionmy{Equivalence of the visual and omniscient model for the alignment term}

Here we show the equivalence between visual and omniscient model for the alignment term  if we consider that each shade is associated to a single particle.

The visual alignment terms is defined as:
\begin{CEquation}\label{eq:S10}
\omega^\mathrm{visu.}_\parallel =
\sum_k  \left( \int^{\varphi_{\Downarrow k}}_{\varphi_{\Uparrow k}} (\mathcal{O}_i(\phi)\cos \phi-\mathcal{D}_i(\phi)\sin \phi) b_\epsilon(\phi)  d\phi\right)\,
\end{CEquation}
 
On the previous section we discussed the piece-wise constant properties of $\mathcal{R}_i(\phi)$, the same is true for $\mathcal{D}_i(\phi)$ and $\mathcal{O}_i(\phi)$, being independent of $\phi$ Taking Eq.~\eqref{eq:S10} and Eq. (4b), we can express the alignment for the visual model as two components, a radial and a azimuthal to be:

\begin{CEquation}\label{eq:S11}
\omega^\mathrm{visu.}_\parallel =\left\langle \frac{\omega_{\parallel,\phi}+\omega_{\parallel,r}}{U}  \right\rangle
\end{CEquation}
\begin{CEquation}\label{azimutal}
\omega_{\parallel,\phi} =
\sum_k \mathcal{O}_k(\Delta\varphi_k) \left( \int^{\varphi_{\Downarrow k}}_{\varphi_{\Uparrow k}}   \cos \phi b_\epsilon(\phi)  d\phi\right)\,
\end{CEquation}
\begin{CEquation}\label{radial}
\omega_{\parallel,r} =
\sum_k -\mathcal{D}_k(\Delta\varphi_k) \left( \int^{\varphi_{\Downarrow k}}_{\varphi_{\Uparrow k}}   \sin \phi b_\epsilon(\phi)  d\phi\right)\,
\end{CEquation}

Eq.~\eqref{radial} is the exact form as Eq. (S2) with the weight of the integral being  $\mathcal{D}_i(\Delta\varphi_k)$ instead of $\mathcal{R}^2_i(\Delta\varphi_k)$ resulting in

\begin{CEquation}\label{radia2l}
\begin{matrix}
    \omega_{\parallel,r} =
\sum_k \frac{-2\mathcal{D}_k(\Delta\varphi_k)}{\mathcal{R}_k(\Delta\varphi_k)}  \sin(\varphi_k)  \left(
        1 + \epsilon\cos\varphi_k\cos\Delta\varphi_k
    \right)
\end{matrix}
\end{CEquation}

As for Eq.~\ref{azimutal} the developing the integral and assuming $\Delta\phi \approx \sin \Delta\phi$ we have:

\begin{CEquation}\label{azimuta2l}
\begin{matrix}
\omega_{\parallel,\phi} =
\sum_k \frac{\mathcal{O}_k(\Delta\varphi_k)}{\mathcal{R}_k(\Delta\varphi_k)}\\
\left(2\cos\varphi_k (1+\epsilon\cos\Delta\varphi_k\cos\varphi_k)+\epsilon(1-\cos\Delta\varphi_k) \right)\,
\end{matrix}
\end{CEquation}

Once more, for the case where there is no occlusion nor aggregation $k=j$ and taking into account the physical constrains of the particles: $\cos\Delta\varphi_k \approx 1$ and $\varphi_k=\theta_{ij}$, then Eqs. (S14-15) becomes 
\begin{CEquation}\label{radia2l}
    \omega_{\parallel,r} \approx\sum_{j\ne i}^N  \frac{-2\mathcal{D}_k(\Delta\varphi_k)}{\mathcal{R}_k(\Delta\varphi_k)} \sin(\theta_{ij})  b_\epsilon(\theta_{ij})
\end{CEquation}
\begin{CEquation}\label{radia2l}
    \omega_{\parallel,\phi} \approx\sum_{j\ne i}^N  \frac{2\mathcal{O}_k(\Delta\varphi_k)}{\mathcal{R}_k(\Delta\varphi_k)} \cos(\theta_{ij})  b_\epsilon(\theta_{ij})
\end{CEquation}

This leads to
\begin{CEquation}\label{eq:S18}
\omega^\mathrm{visu.}_\parallel \approx \left\langle \sum_{j\ne i}^N  2\frac{\boldsymbol{e}_i \times \boldsymbol{V}_{ij}}{\mathcal{R}^2_k(\Delta\varphi_k)}  b_\epsilon(\theta_{ij}) \right\rangle
\end{CEquation}

Now $\boldsymbol{V}_{ij}=U(\boldsymbol{e}_j-\boldsymbol{e}_i)$ then $\boldsymbol{e}_i \times \boldsymbol{e}_j = \frac{\boldsymbol{e}_i\times \boldsymbol{V}_{ij}}{U}$. And $\mathcal{R}_i(\Delta\varphi_k)=||\boldsymbol{x}_j-\boldsymbol{x}_i||$, resulting in

\begin{CEquation}\label{eq:S19}
\omega^\mathrm{visu.}_\parallel \approx \left\langle \sum_{j\ne i}^N  2\frac{\boldsymbol{e}_i \times \boldsymbol{e}_{j}}{||\boldsymbol{x}_j-\boldsymbol{x}_i||^2}  b_\epsilon(\theta_{ij}) \right\rangle
\end{CEquation}

The normalization nullifies the factor 2 that is not present in the omniscient model and shows that the two models are equivalent when shades are formed by single particles.
\newpage
\subsectionmy{Blind spot} 

\begin{figure}
  \centering
  \includegraphics[width=0.9\columnwidth]{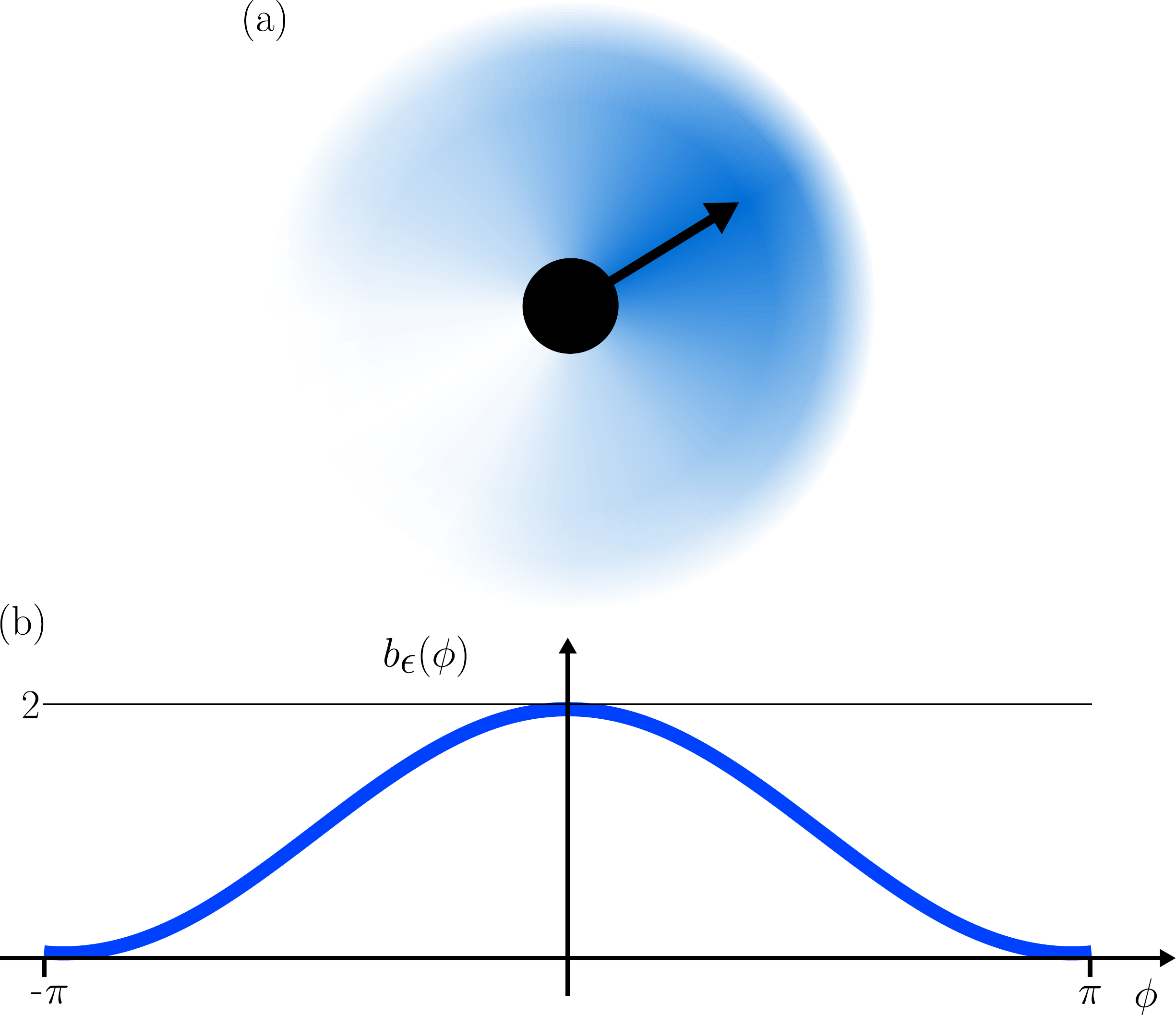}
  \caption{Effect of a blind spot. (a) 2D view: When $\epsilon = 1$, each particle cannot see at its rear (transparent), but the view is normal in front (blue). (b) Function $b_\epsilon(\phi)=1+\epsilon \cos\phi$ when $\epsilon = 1$}
  \label{supfig:blindspot}
\end{figure}

Figure Fig. S\ref{supfig:blindspot} illustrates the function $b_\epsilon(\varphi)$ used to model a blind spot in the visual field. When $\epsilon$ is equal to 0 there is no effect and with 1 the effect is maximum.

\sectionmy{Digital implementation}

\subsectionmy{Omniscient model}

As previously stated, the model that we propose has an omniscient implementation.
The alignment term is shown on Eq.~\eqref{eq:S7} can be normalized by Eq. (5), which lead to the implementation of:

\begin{CEquation}
    \omega_\parallel=\left\langle \sum^N_{j\ne i} \frac{\boldsymbol{e}_i \cross \boldsymbol{e}_j}{||\boldsymbol{x}_j-\boldsymbol{x}_i||^2 } b_\epsilon(\theta_{ij})\right\rangle
\end{CEquation}

For attraction term, the implemented equation is:

\begin{CEquation}
    \label{eq:OmniAtt}
    \omega_\odot=\left\langle \sum^N_{j\ne i} ||\boldsymbol{x}_j-\boldsymbol{x}_i||\sin{ (\theta_{ij})}b_\epsilon(\theta_{ij})\right\rangle
\end{CEquation}
\subsectionmy{Visual model}

On the main text, we described the continuous form of the equations for attraction (Eq. (4a)) and for alignment (Eq. (4b)) and Sections S1.1 and S1.3 expanded on equivalency between these visual terms when there is no occlusion or aggregation.  Here, we go in more depth on how the visual components are derived and calculated, as well as the formulation of the implementation for equations for the numerical simulations.

Consider the point of view of the i-th particle in 2 consecutive time steps for the j-th particle:
\begin{figure}[t]
  \centering
  \includegraphics[width=0.6\columnwidth]{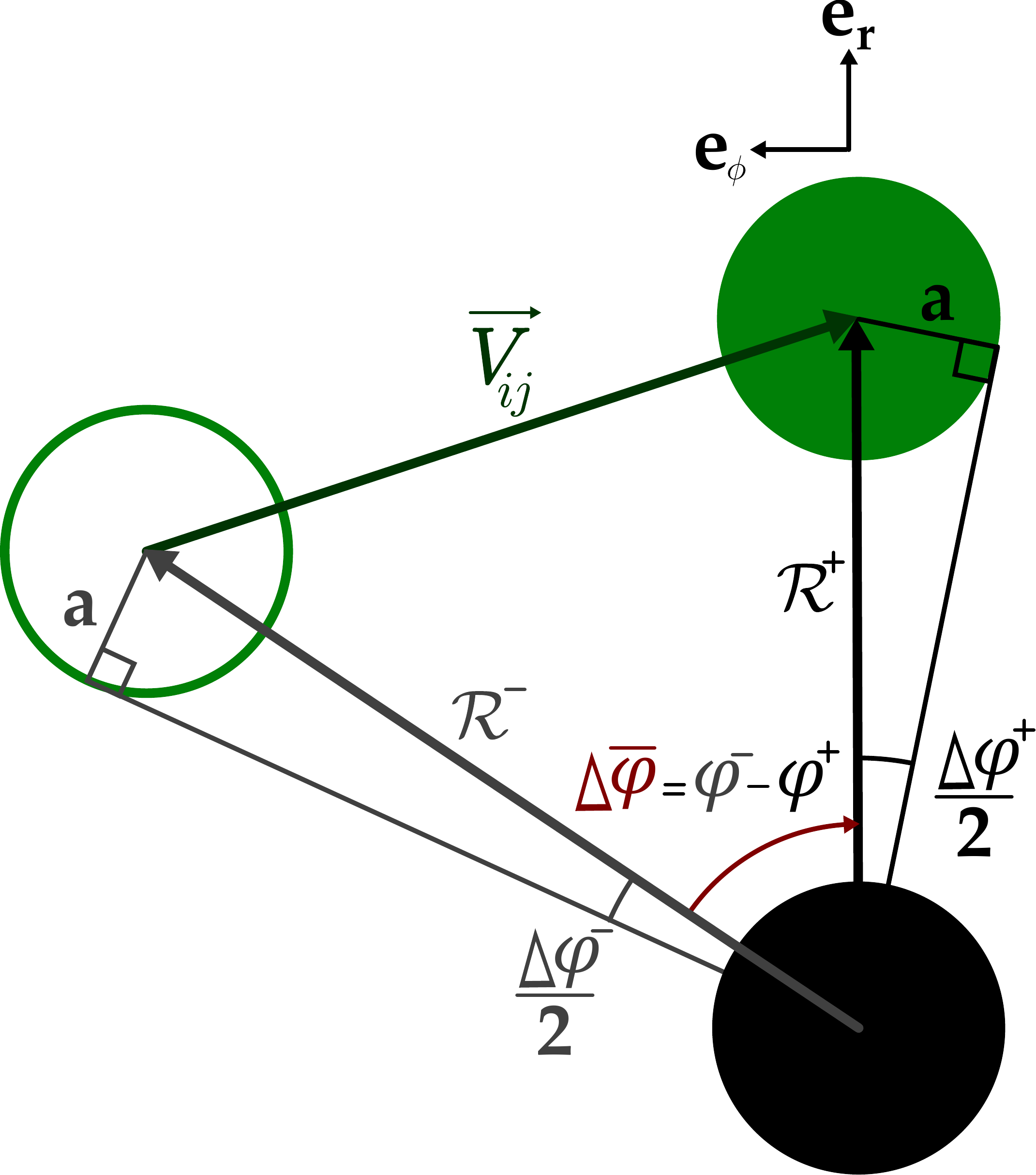}
  \caption{Point of view of the i-th particle (black) of the j-th particle in t-1 (contour) and t (solid color)}
\label{fig:phases}
\end{figure}

Each particle has a radius $a$, the over index represent + time t and - time t-dt. $\boldsymbol{e}_r$ and  $\boldsymbol{e}_\phi$ are the unit vectors of the polar coordinates referred to $\theta_i$, where the former is radial and the later azimuthal. $\overrightarrow{V}_{ij}$ is the relative velocity vector perceived by the particle i of the net velocity of the particle j. Then from the geometry of the problem we have
\begin{CEquation}\label{eq:S20}
    \mathcal{R}_i^{\pm}=\frac{a}{\sin\Delta\varphi^\pm}
\end{CEquation}
Then the $\mathcal{R}_i$ vectors can be described as follows:
\begin{CEquation}\label{eq:S21}
    \overrightarrow{\mathcal{R}_i^+}=\mathcal{R}_i^+ \boldsymbol{e}_r 
\end{CEquation}
\begin{CEquation}\label{eq:S22}
    \overrightarrow{\mathcal{R}_i^-}=\mathcal{R}_i^-[\cos{\Delta\bar\varphi}\boldsymbol{e}_r -\sin{\Delta\bar\varphi}\boldsymbol{e}_\phi] 
\end{CEquation}
Consequently the distance associated with the movement of  $\overrightarrow{V}_{ij}$ is:
\begin{CEquation}\label{eq:S23}
    U\overrightarrow{V}_{ij}\delta t=(\mathcal{R}_i^+-\mathcal{R}_i^-\cos{\Delta\bar\varphi})\boldsymbol{e}_r +\mathcal{R}_i^-\sin {\Delta\bar\varphi}\boldsymbol{e}_\phi
\end{CEquation}
\begin{CEquation}\label{eq:S24}
    \overrightarrow{V}_{ij}=\frac{[\mathcal{R}_i^+-\mathcal{R}_i^-\cos{\Delta\bar\varphi},\mathcal{R}_i^-\sin {\Delta\bar\varphi}]}{U\delta t}
\end{CEquation}
From Eq.~\eqref{eq:S24} we can do the decomposition in radial and azimutal components to calculate $\mathcal{O}_i$ and $\mathcal{D}_i$
\begin{CEquation}
    \mathcal{O}_i=\frac{1}{U\delta t}-\frac{\mathcal{R}_i^-\cos{\Delta\bar\varphi}}{\mathcal{R}_i^+U\delta t}\label{eq:Vper}
\end{CEquation}

\begin{CEquation}
    \mathcal{D}_i=\frac{\mathcal{R}_i^-\sin {\Delta\bar\varphi}}{\mathcal{R}_i^+U\delta t}\label{eq:Vpar}
\end{CEquation}
Note that on a robotic implementation this method of calculating the optic flow and its divergence ($\mathcal{O}_i$ and $\mathcal{D}_i$), would be entirely replaced by its direct measurement via a specialized optic flow sensor.

For the computer implementation of the alignment term of the visual model, we use Eq.~\eqref{eq:S24} to calculate the cross product of the relative vectors over $\mathcal{R}_i^2(\Delta \varphi_k)$

\begin{CEquation}\label{eq:S27}
    \omega_\parallel=\left\langle\sum_k \sin^2(\Delta\varphi_k)\frac{ \overrightarrow{V}_{ik}\cross \boldsymbol{e}_i}{Ua^2}  \left(
        1 + \epsilon\cos\varphi_k
    \right)\right\rangle
\end{CEquation}
 Note that there are two approximations ($\cos\Delta\varphi_k \approx 1$ and $\sin\Delta\varphi_k \approx \Delta\varphi_k $) on Eq.~\eqref{eq:S27}.

As for attraction, we use at Eq.~\eqref{eq:S6} written on the terms described above.
\begin{CEquation}
    \label{eq:attstep1}
    \omega_\odot=\left\langle\sum_k 2a^2\frac{ \sin(\varphi_k)}{\sin(\Delta\varphi_k)}  \left(
        1 + \epsilon\cos\varphi_k\cos\Delta\varphi_k
    \right)\right\rangle
\end{CEquation}

\subsectionmy{Algorithm to detect corresponding features necessary to compute optic flow}

The optic flow is the perception of relative movement and we use a feature matching-inspired algorithm to calculate it in numeric simulation. But this feature match requires that you discriminate and detect the same feature in two consecutive time steps. Here we present the algorithm to match different features given the binary characteristic of $\mathcal{V}_i(\phi)$. In the model this will be the assumption to use Eqs.~\eqref{eq:Vper} and \eqref{eq:Vpar}. Note that shades can split or join in between time steps. To detect these events, we employ a minimum distance error to correlate two set of features in two consecutive time steps. 

\begin{figure}
  \centering
  \includegraphics[width=1\columnwidth]{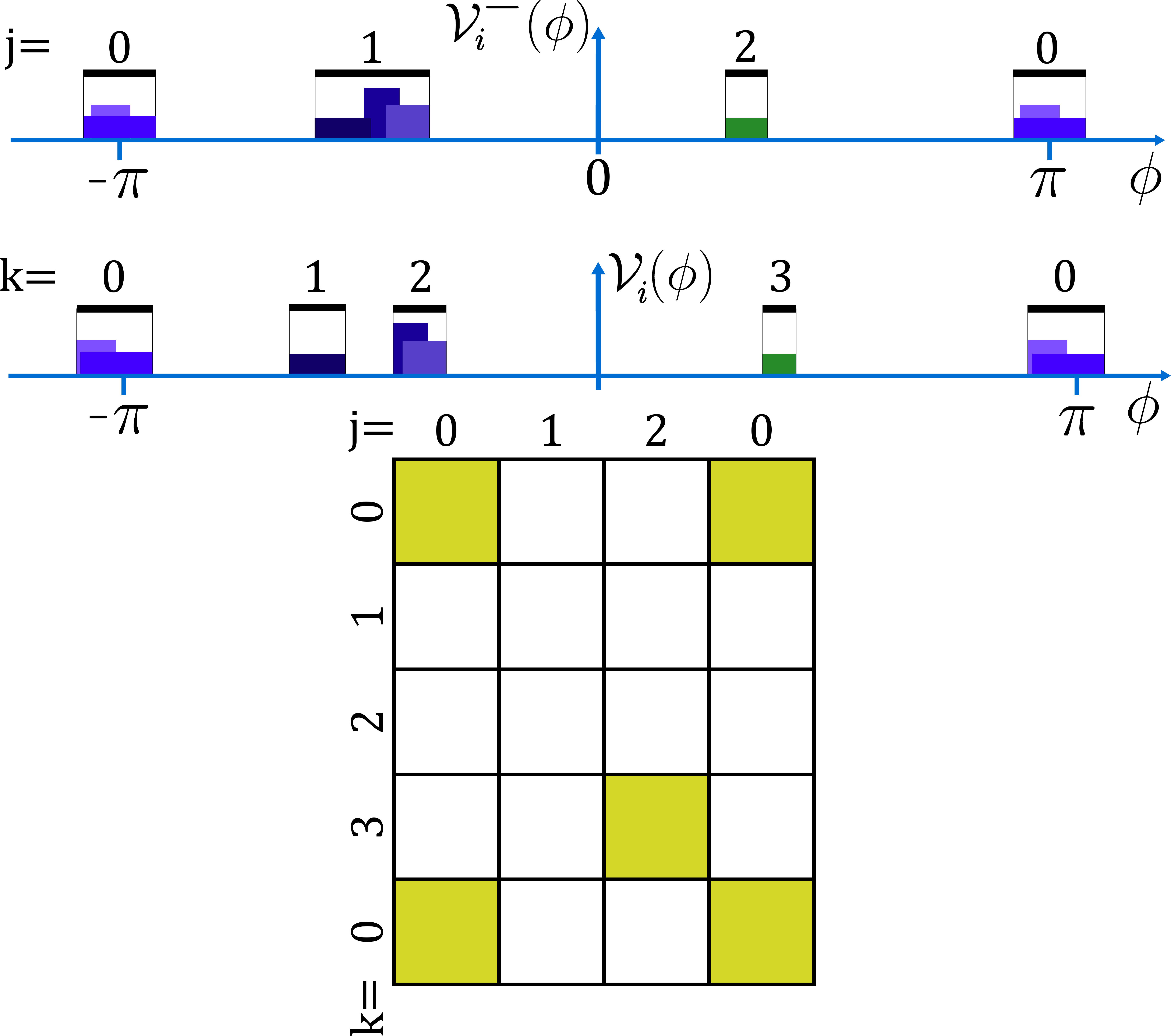}
  \caption{Example of two consecutive visual fields. The matrix at the bottom illustrate the algorithm we use to find corresponding shades between two time steps.  $\Delta\bar{\phi}$ is the delta between the mid point of the shades between two consecutive time steps.}
  \label{supfig:ofalgo}
\end{figure}

In  Fig. S\ref{supfig:ofalgo}, we illustrate the  algorithm used, where yellow boxes are associated to corresponding shades. To correspond to each other, shades should be associated to radial and azimuthal velocities bounded by $U$. 

This algorithm is also illustrated on the following pseudo-code (see Algorithm \ref{alg:cap}).
\begin{figure}
\begin{algorithmic}
\State $\overline{\mathcal{V}_{i_{t-1}}} \gets [(\varphi_{\Uparrow l}+\varphi_{\Downarrow l})/2]$ at t-1 $\forall \varphi$ in $\phi_{i,t-1}$
\State $\Delta\mathcal{V}_{i_{t-1}} \gets [\varphi_{\Downarrow l}-\varphi_{\Uparrow l}]$ at t-1 $\forall \varphi$ in  $\phi_{i,t-1}$
\State $\overline{\mathcal{V}_{i_{t}}} \gets [(\varphi_{\Uparrow k}+\varphi_{\Downarrow k})/2]$ at t $\forall \varphi$ in $\phi_{i,t}$
\State $\Delta\mathcal{V}_{i_{t}} \gets [\varphi_{\Downarrow k}-\varphi_{\Uparrow k}]$ at t $\forall \varphi$ in $\phi_{i,t}$
\State $\omega \gets \omega_{k,t-1} \delta t$ \Comment{Previous known rotation}
\Require $l \& k \geq 0$
\State $VisualDist \gets \|\overline{\mathcal{V}_{i_{t-1}}}-\overline{\mathcal{V}_{i_{t}}}\|$ \Comment{A k x l matrix}
\State $RefDist \gets min(\sin{\Delta\mathcal{V}_{i_{t}}},\sin{\Delta\mathcal{V}_{i_{t-1}}}) 2U\delta t/a+\omega$
\State $VisualArea \gets \| \frac{1}{\Delta\mathcal{V}_{i_{t}}}-\frac{1}{\sin{\Delta\mathcal{V}_{i_{t-1}}}}\|$ \Comment{A k x l matrix}
\State $RefArea \gets 2U\delta t/a$
\For{all values in $VisualArea,VisualDist,RefDist \And RefArea$ }
\If{$ ValueVisualArea $ < $ RefDist  \And  ValueVisualArea $ < $ ValueVisualArea $}
    \State $Pair_{t} \gets [\varphi_{\Uparrow k},\varphi_{\Downarrow k}]$
    \State $Pair_{t-1} \gets [\varphi_{\Uparrow l},\varphi_{\Downarrow l}]$
\EndIf
\EndFor

\end{algorithmic}
\par Corresponding shades between time steps\label{alg:cap}
\end{figure}

The model is implemented on python 3.10 (it will be available for download on \cite{GitLab} and \cite{osf}. This includes the implementation of the algorithm described by Fig. S\ref{supfig:ofalgo}, all model equations, data management and figure creation. There is a markdown file that only include the model implementation without data logging. 

\newpage
\begin{widetext}
\sectionmy{Detailed Phase Diagrams (Fig. 3)}
\subsectionmy{Visual Model} 

\begin{figure}[hb]
  \centering
  \includegraphics[width=1\columnwidth]{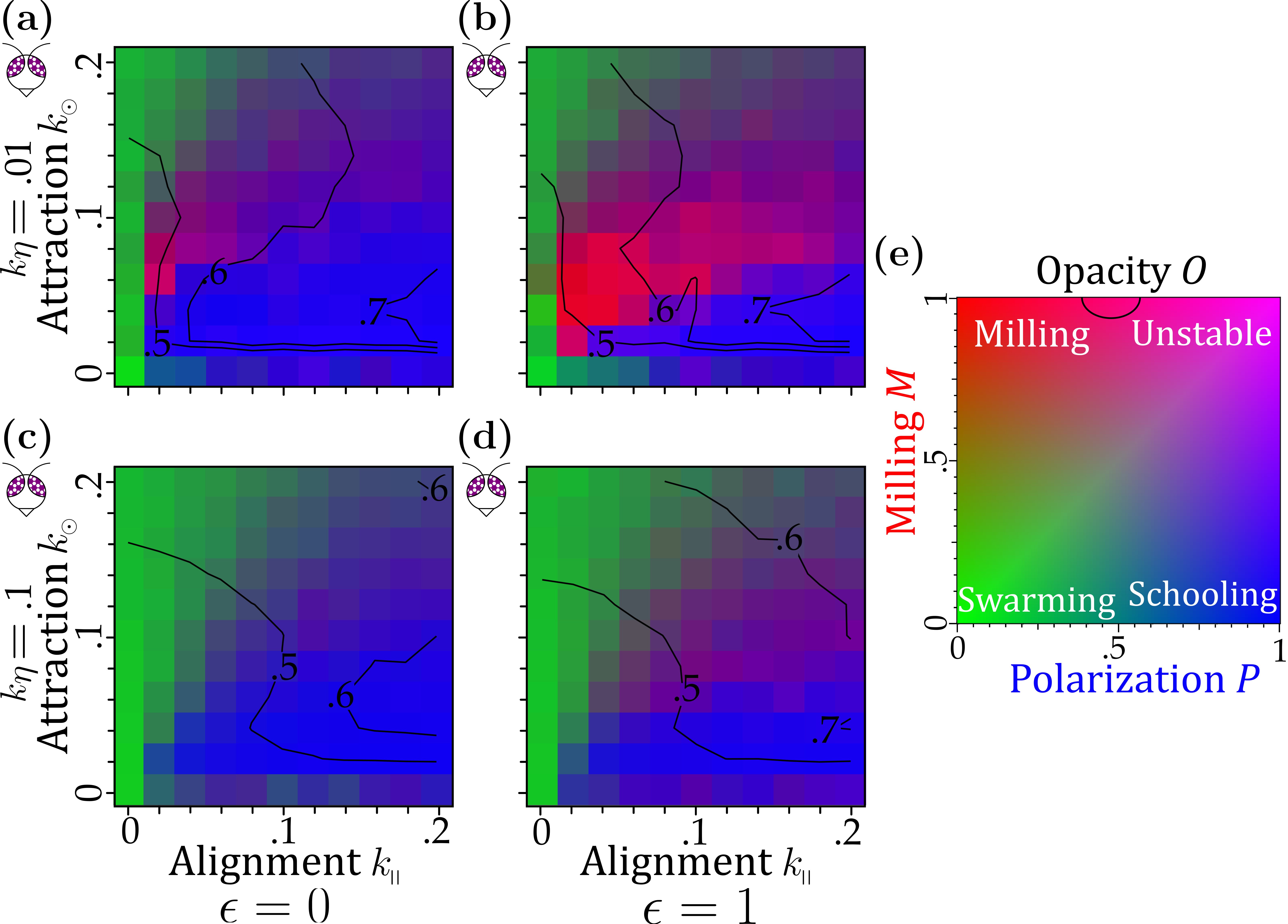}
  \caption{Phase diagrams for $N=50$ individuals. These phase diagrams are for: (a)-(c) noise strength $k_\eta=0.01$, where there is no blind angle ($\epsilon =0$) and maximum ($\epsilon =1$), respectively. (b)-(d) noise strength $k_\eta=0.1$, where there is no blind angle ($\epsilon =0$) and maximum ($\epsilon =1$), respectively  In (e), is illustrated how, in the phase space $(k_\parallel, k_\odot)$, the colors represents different values of $P$ and $M$, and contours show the opacity $O$; these colors also represents the phases discussed earlier.}
\end{figure}
\begin{figure}
  \centering
  \includegraphics[width=\textwidth]{SFig5.pdf}
  \caption{Phase Diagram and Phase Reproduction for different Flock sizes for the Visual model. i) Phase Diagram, ii)Swarming, iii)Schooling, iv)Milling. a)N=5, b)N=10, c)N=30,  d)N=100, e)N=300, }
\end{figure}
\begin{figure}
  \centering
  \includegraphics[width=0.9\textwidth]{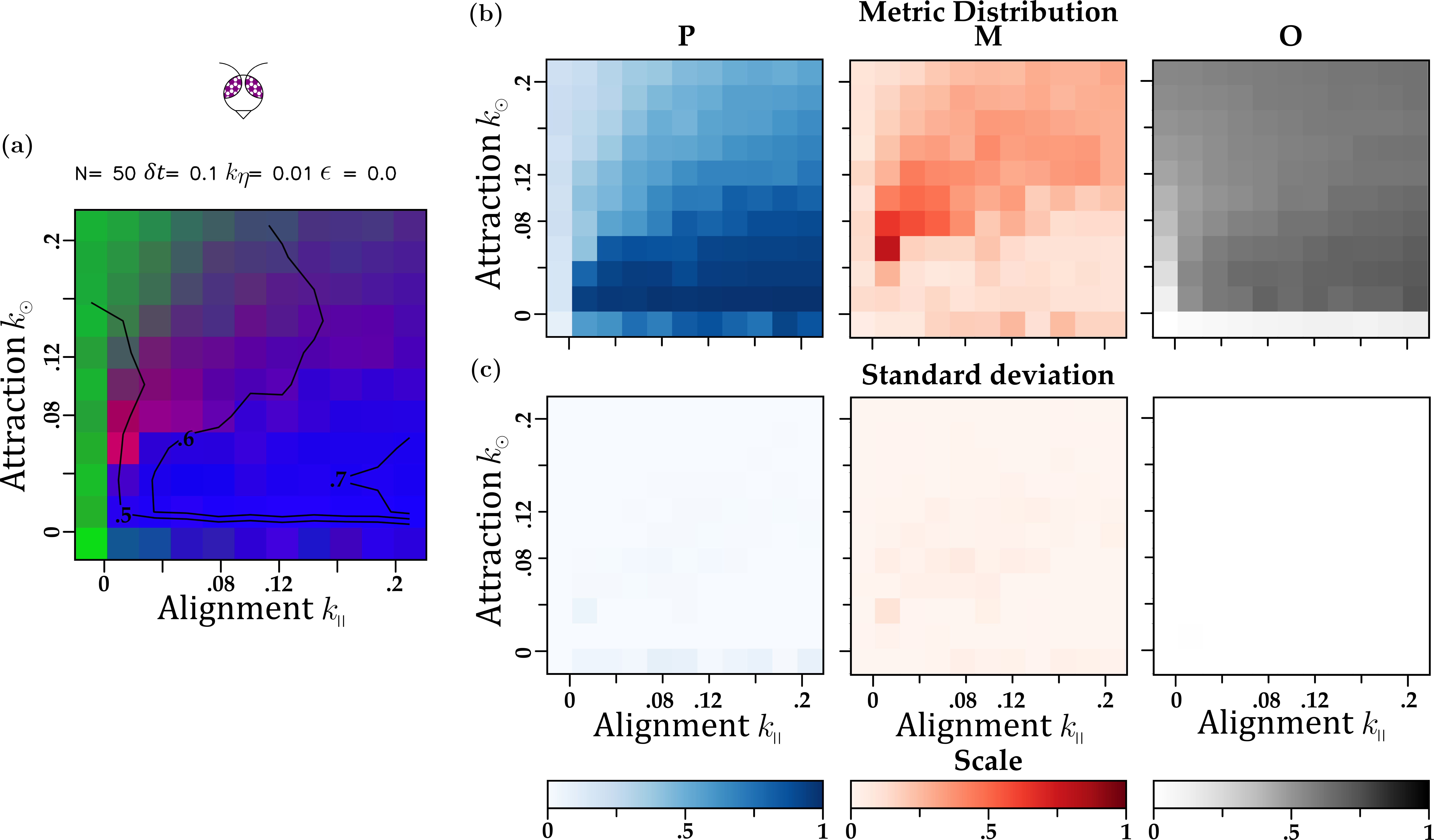}
  \caption{\variancecaption{Visual}{50}{0.01}{no}}
\end{figure}
\begin{figure}
  \centering
  \includegraphics[width=0.9\textwidth]{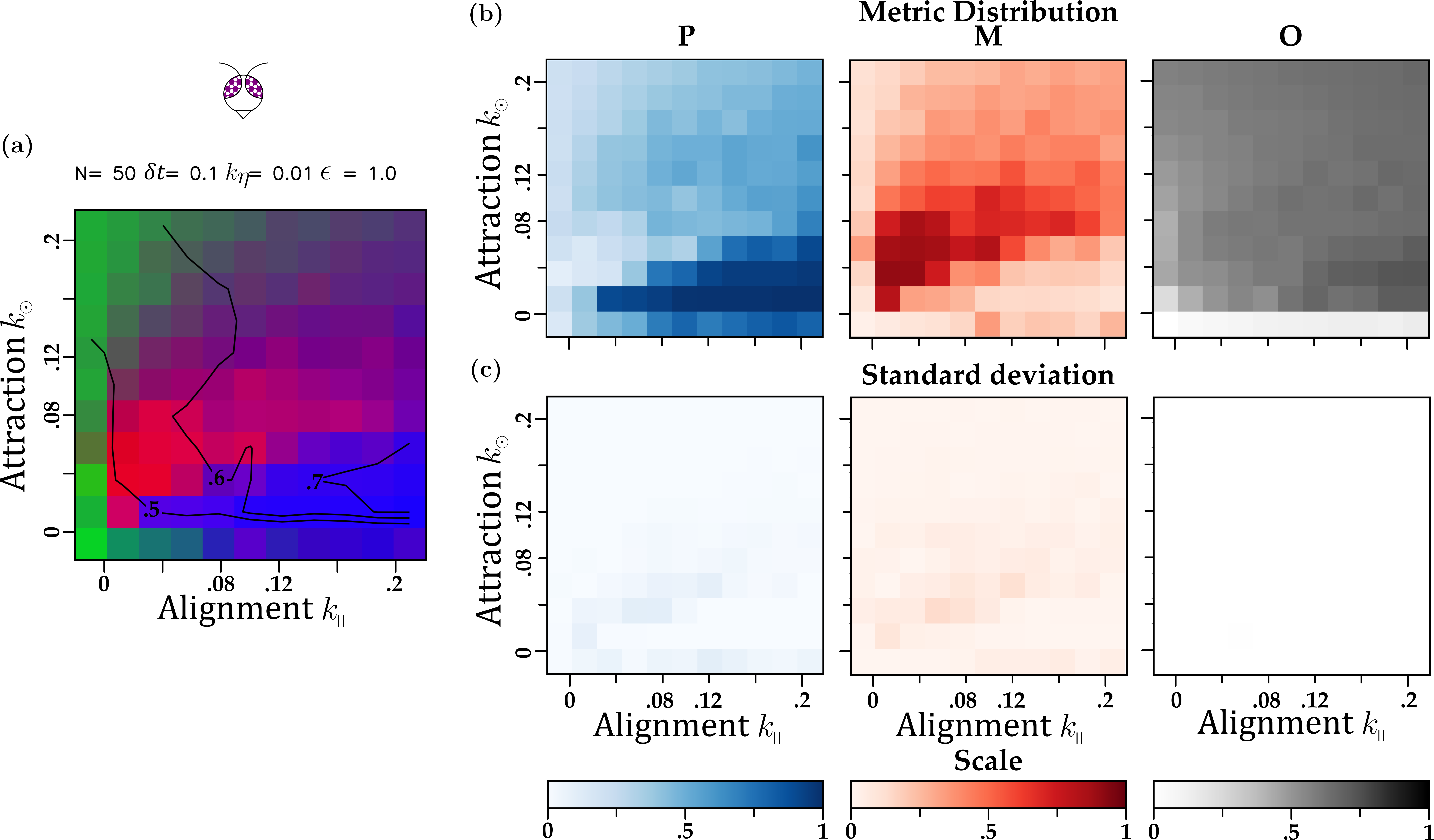}
  \caption{\variancecaption{Visual}{50}{0.01}{maximum}}
\end{figure}
\begin{figure}
  \centering
  \includegraphics[width=0.9\textwidth]{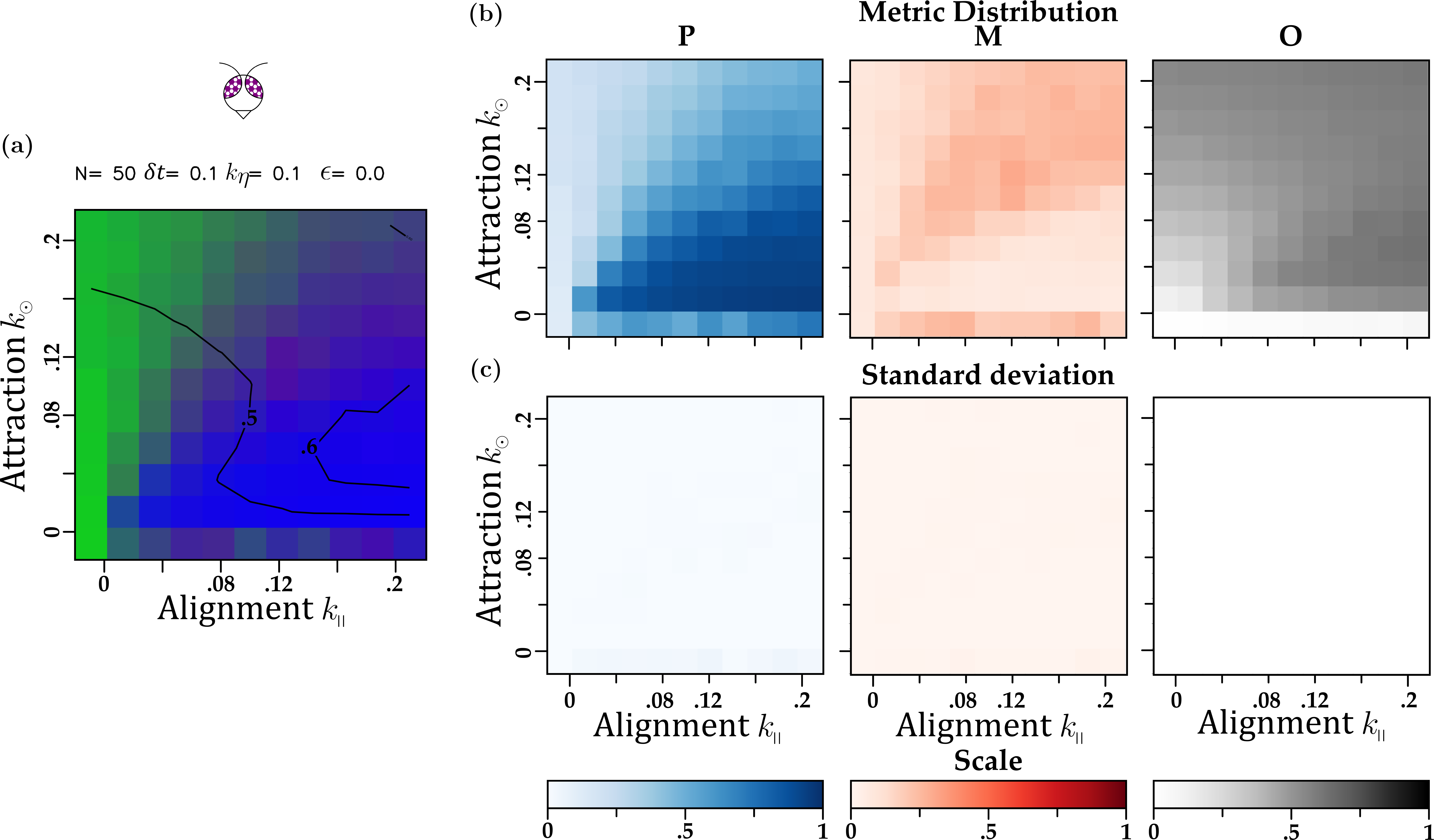}
  \caption{\variancecaption{Visual}{50}{0.1}{no}}
\end{figure}
\begin{figure}
  \centering
  \includegraphics[width=0.9\textwidth]{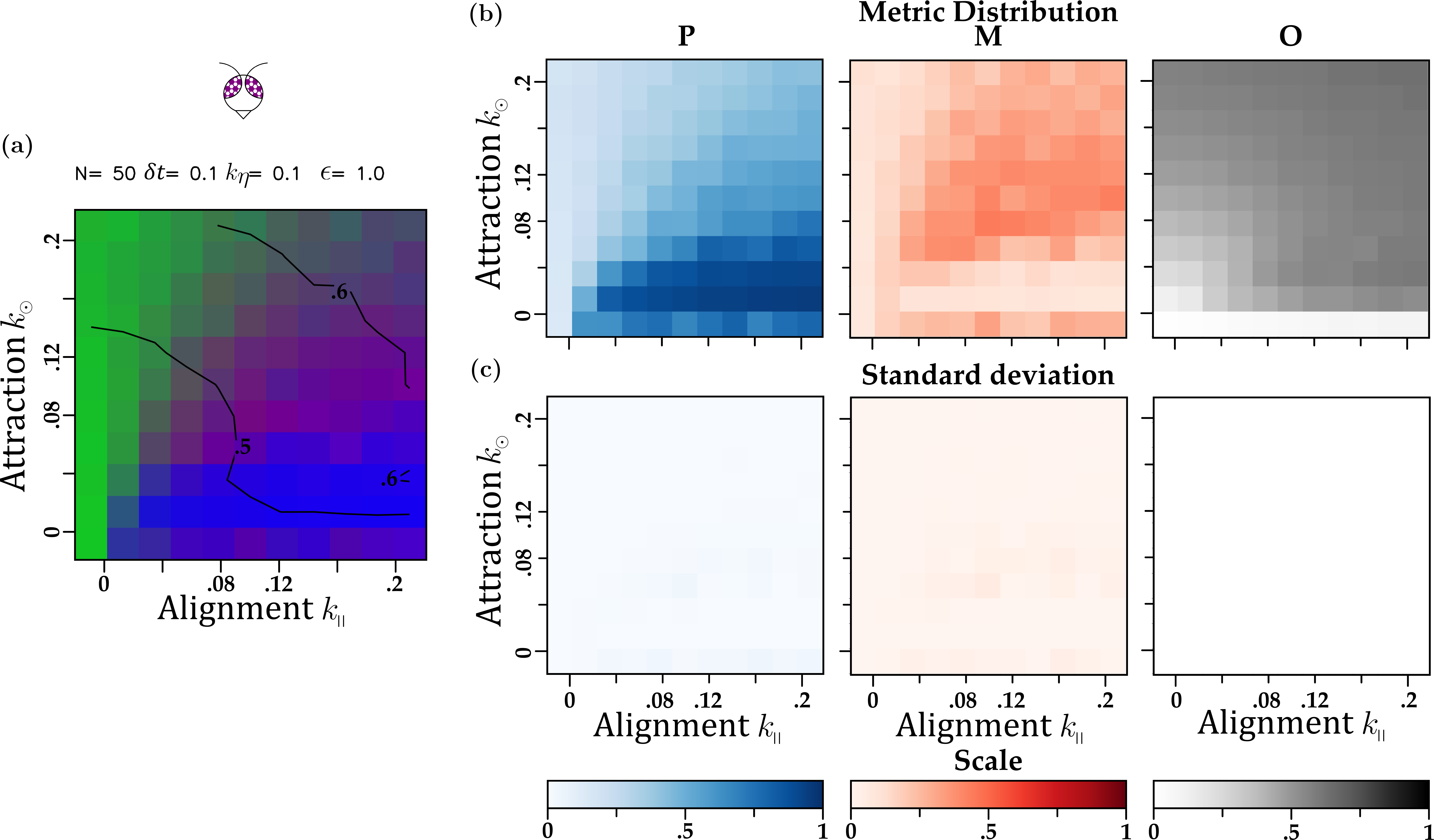}
  \caption{\variancecaption{Visual}{50}{0.1}{maximum}}
\end{figure}

\end{widetext}
\begin{widetext}
\newpage

\subsectionmy{Omniscient Model} 
\begin{figure}[ht]
  \centering
  \includegraphics[width=0.9\columnwidth]{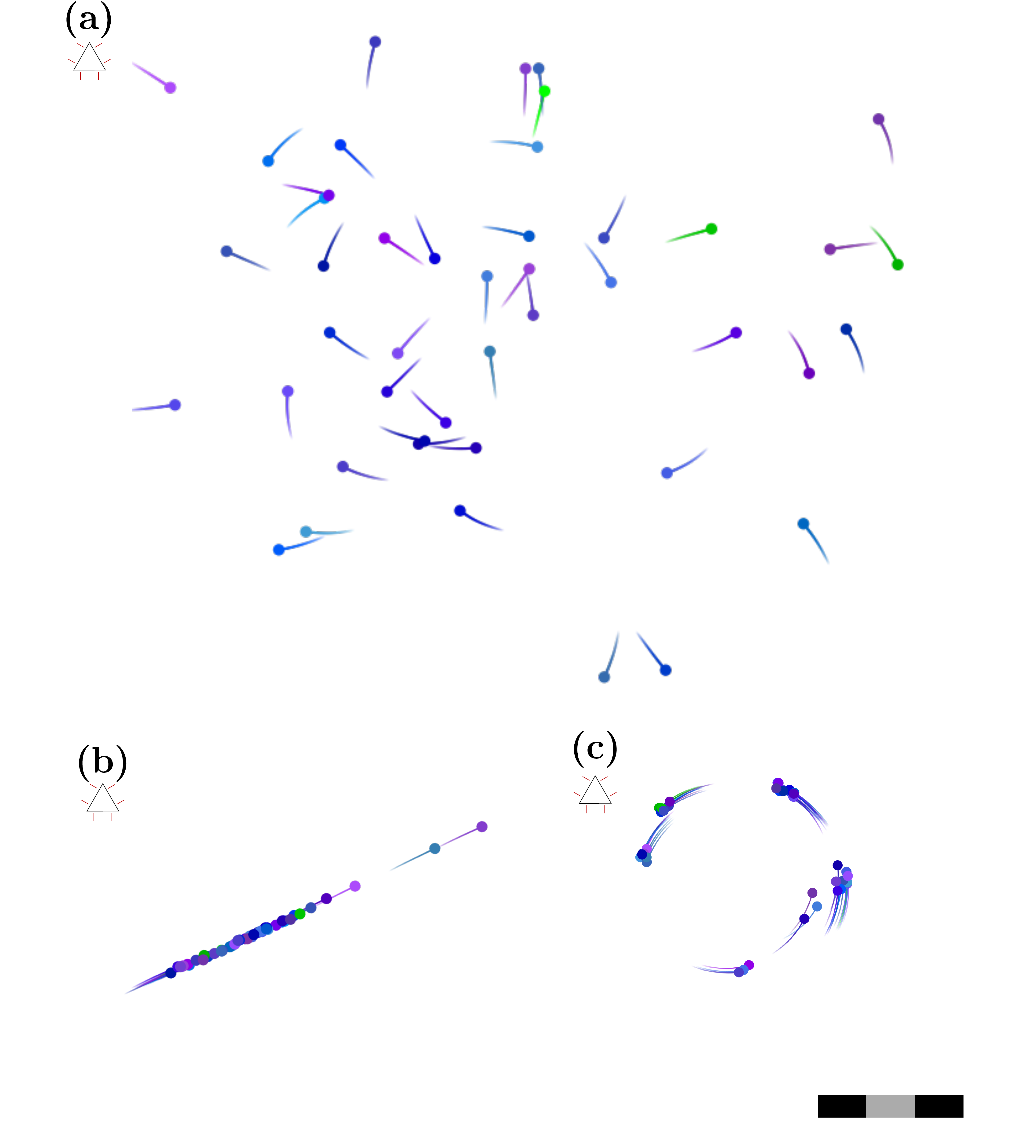}
  \caption{Illustration of the phases observed for the omniscient model and a flock size of $N=50$ individuals. With  $k_\eta=0.01$. The scale bar measures $30a$ ($10a$ each section). We observe three phases:
  (a) Swarming ($k_\odot=0.06$, $k_\parallel=0,\epsilon=0 $);
 (b) Schooling ($k_\odot=0.06$, $k_\parallel=0.2,\epsilon=1$);
  (c)Milling ($k_\odot=0.06$, $k_\parallel=0.06,\epsilon=1$).}
\end{figure}
\begin{figure}[hb]
  \centering
  \includegraphics[width=1\columnwidth]{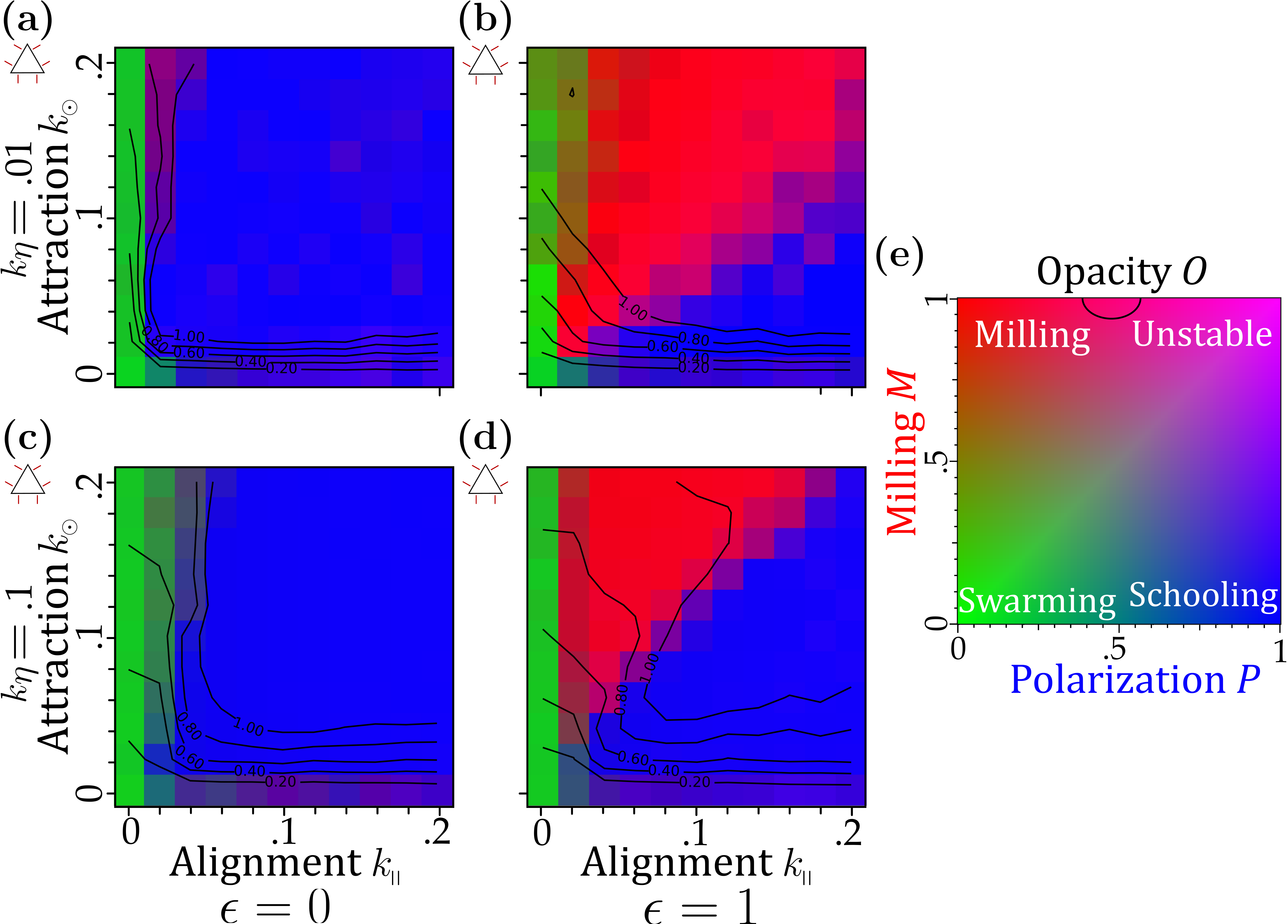}
  \caption{Phase diagrams for omniscient model and flock size of $N=50$ individuals. These phase diagrams are for: (a)-(c) noise strength $k_\eta=0.01,\epsilon =0$ and $\epsilon =1$, respectively. (b)-(d) noise strength $k_\eta=0.1,\epsilon =0$ and $\epsilon =1$) respectively  In (e), is illustrated how, in the phase space $(k_\parallel, k_\odot)$, the colors represents different values of $P$ and $M$, and contours show the opacity $O$; these colors also represents the phases discussed earlier.}
\end{figure}
\newpage
\begin{figure} [h]
  \centering
  \includegraphics[width=0.95\textwidth]{SFig21.pdf}
  \caption{Phase Diagram and Phase Reproduction for different Flock sizes for the Visual model. i) Phase Diagram, ii)Swarming, iii)Schooling, iv)Milling. a)N=5, b)N=10, c)N=30,  d)N=100, e)N=300.}
\end{figure}
\begin{figure} [h]
  \centering
  \includegraphics[width=0.85\textwidth]{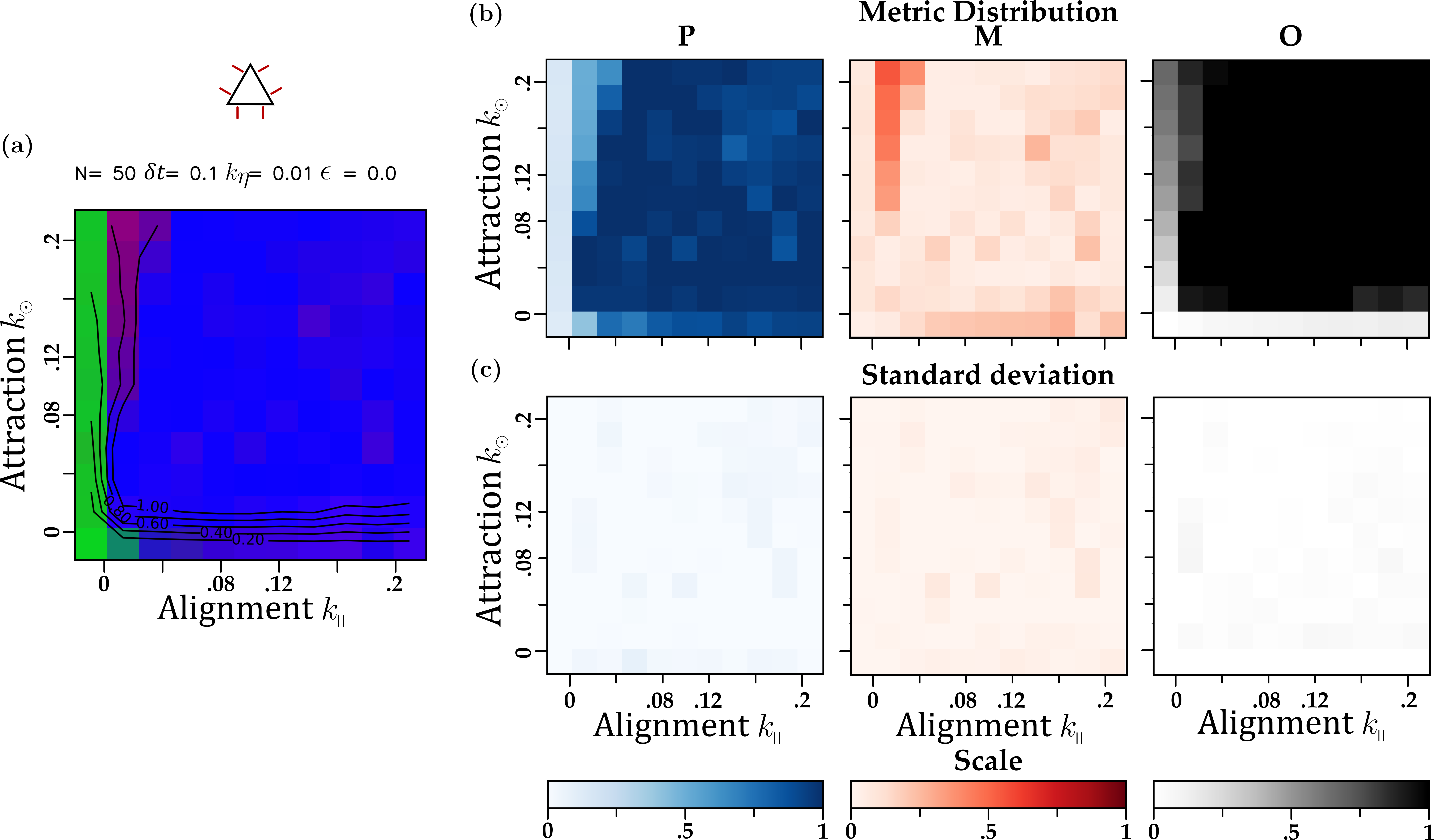}
  \caption{\variancecaption{Omniscient}{50}{0.01}{no}}
\end{figure} 
\begin{figure} [h]
  \centering
  \includegraphics[width=0.85\textwidth]{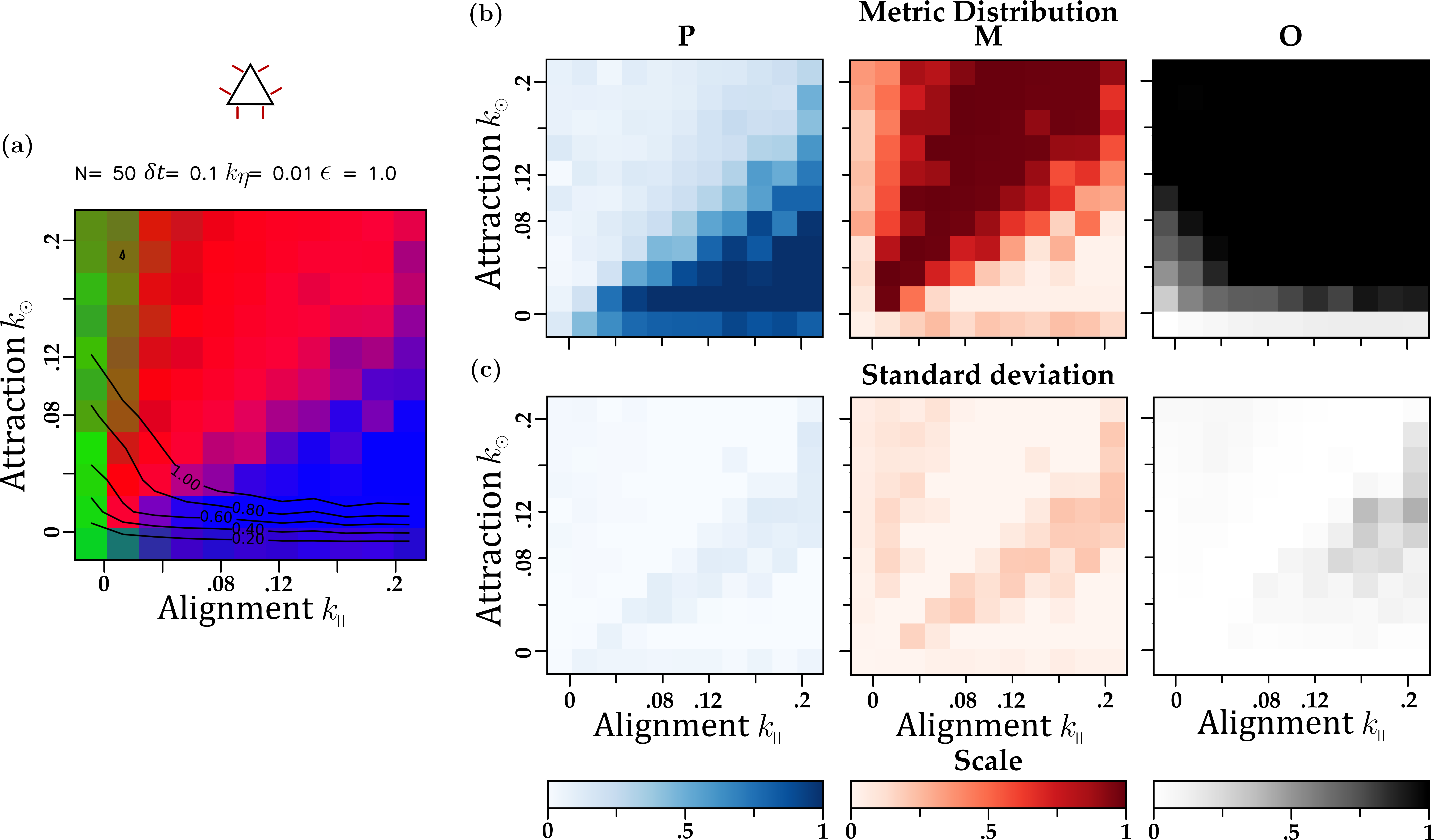}
  \caption{\variancecaption{Omniscient}{50}{0.01}{maximum}}
\end{figure} 
\begin{figure}[h]
  \centering
  \includegraphics[width=0.85\textwidth]{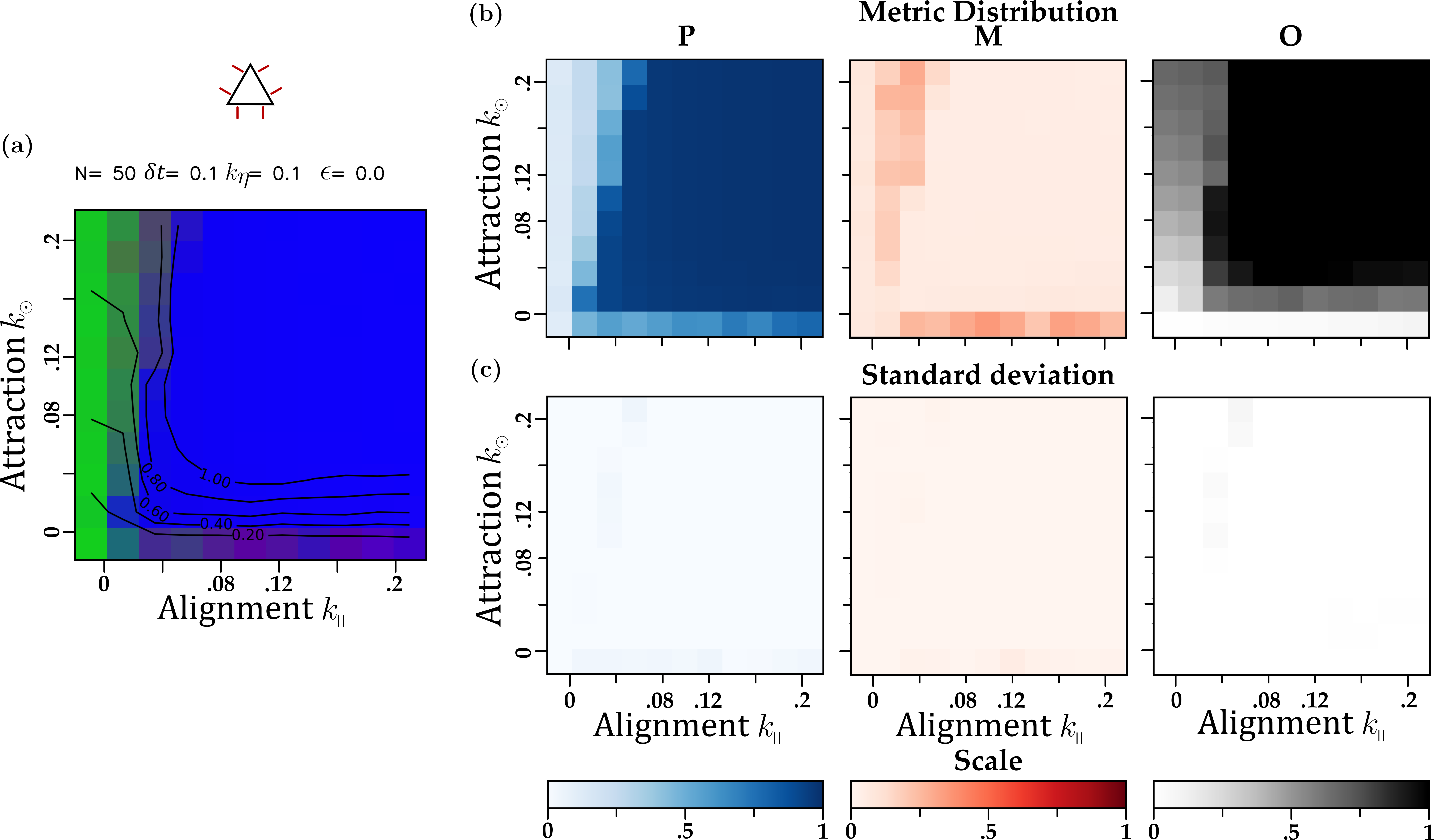}
  \caption{\variancecaption{Omniscient}{50}{0.1}{no}}
\end{figure}
\begin{figure}[h]
  \centering
  \includegraphics[width=0.85\textwidth]{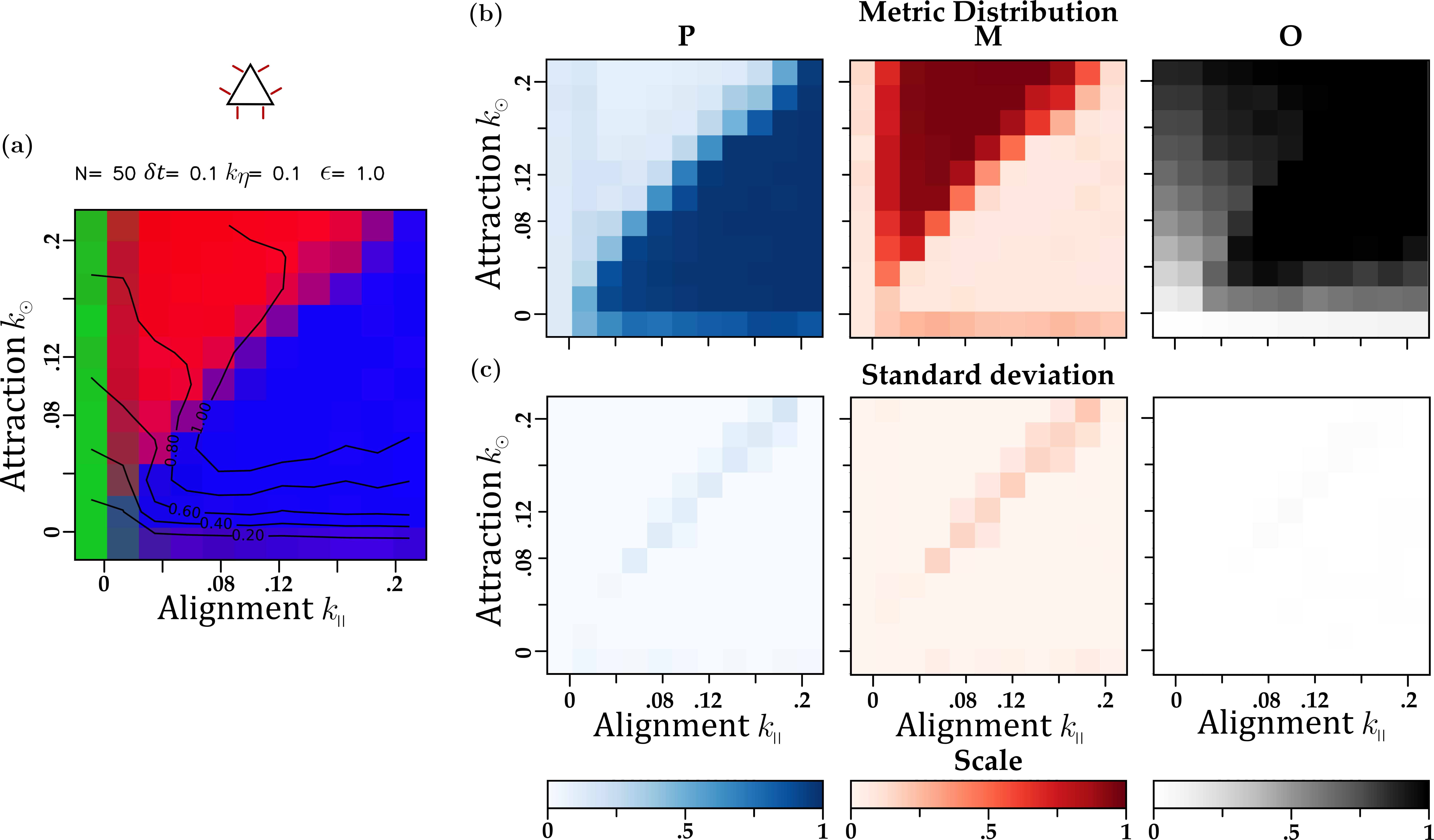}
  \caption{\variancecaption{Omniscient}{50}{0.1}{maximum}}
\end{figure}
\end{widetext}
\clearpage
\begin{widetext}
\sectionmy{Video Snapshots of Time step stability analysis}
\subsectionmy{Visual model}
\begin{figure}[hb]
  \centering
  \includegraphics[width=0.85\textwidth]{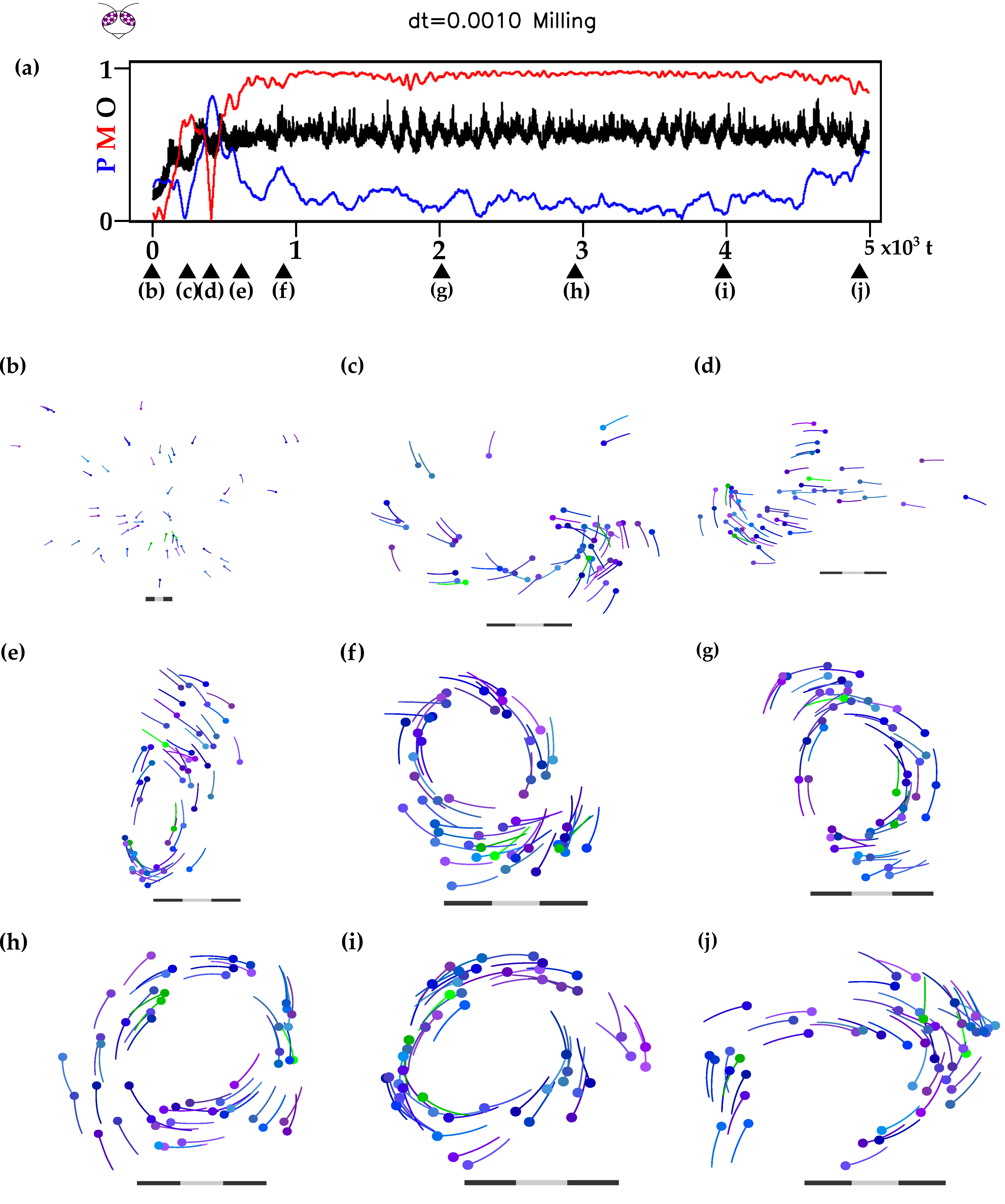}
  \caption{\dtcaption{0.001}{Visual}, Video S1 (\url{https://youtu.be/DDjb5NUwKIU}).}
\end{figure}
\begin{figure}[hb]
  \centering
  \includegraphics[width=1\textwidth]{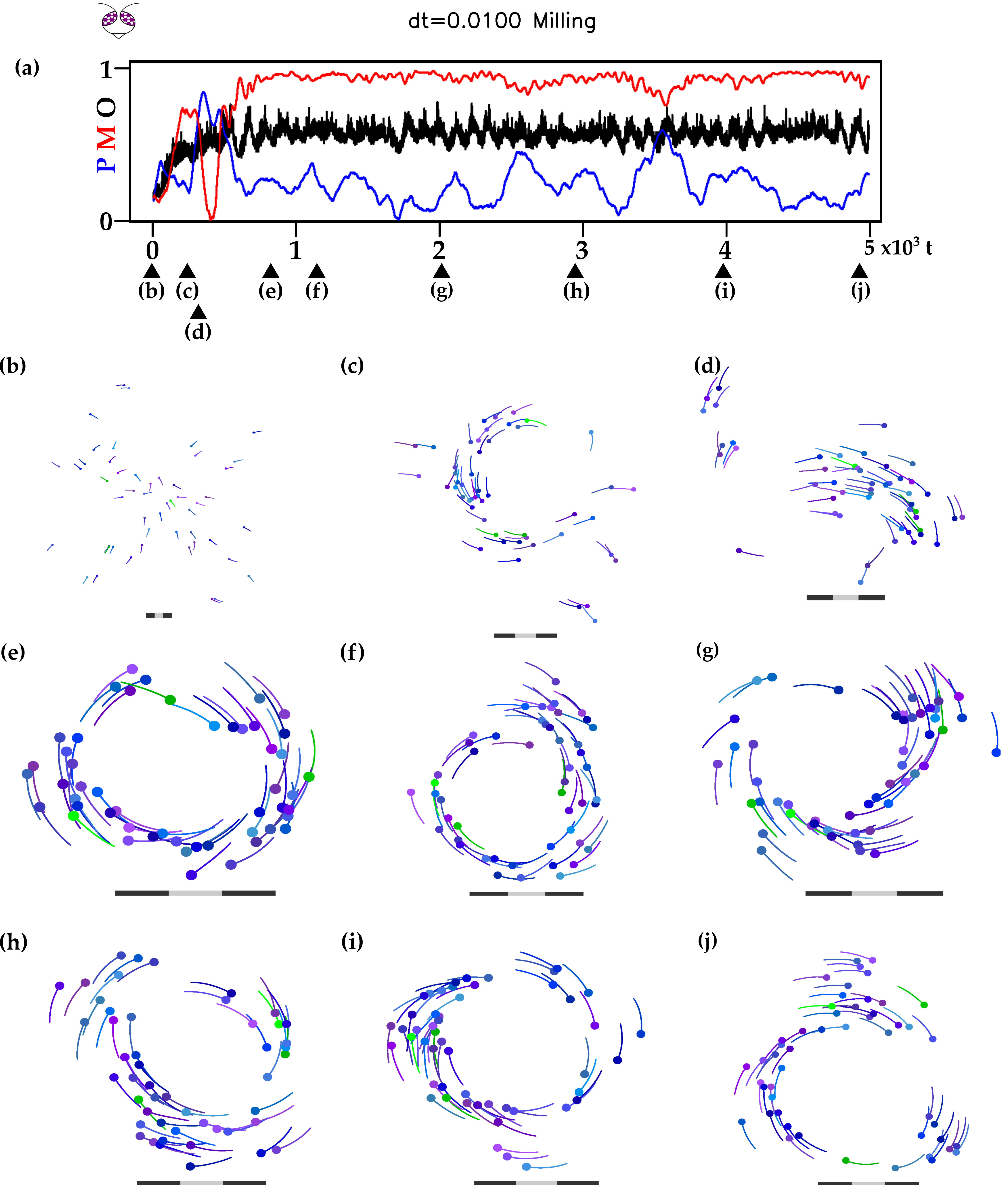}
  \caption{\dtcaption{0.01}{Visual}, Video S2 (\url{https://youtube.com/shorts/dOo7oabdQKo}).}
\end{figure}
\begin{figure}[hb]
  \centering
  \includegraphics[width=1\textwidth]{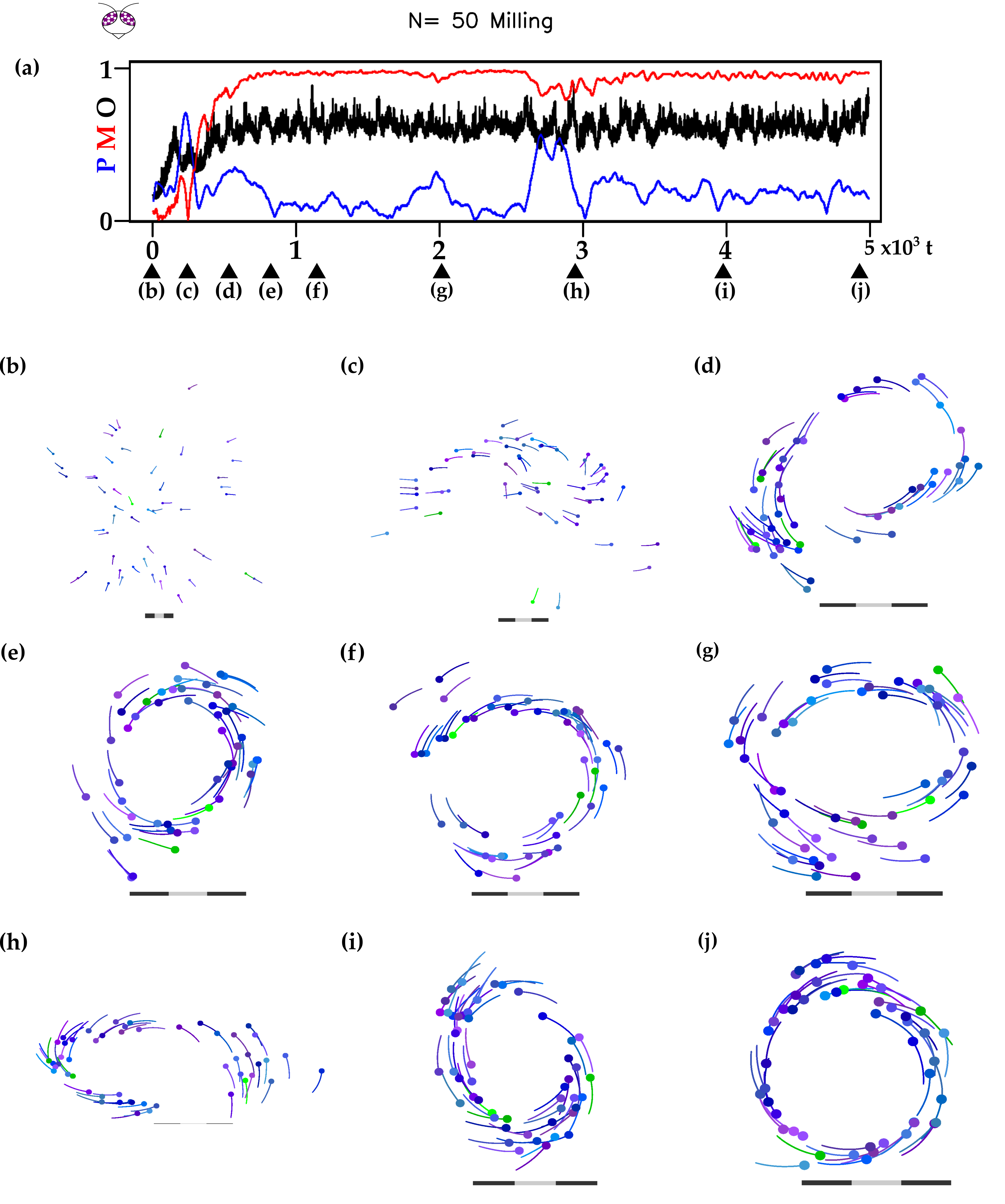}
  \caption{\dtcaption{0.1}{Visual}, Video S3 (\url{https://youtu.be/1KMUSWcKyN8}).}
\end{figure}
\end{widetext}

\begin{widetext}
\clearpage
\subsectionmy{Omniscient model}
\begin{figure}[hb]
  \centering
  \includegraphics[width=0.85\textwidth]{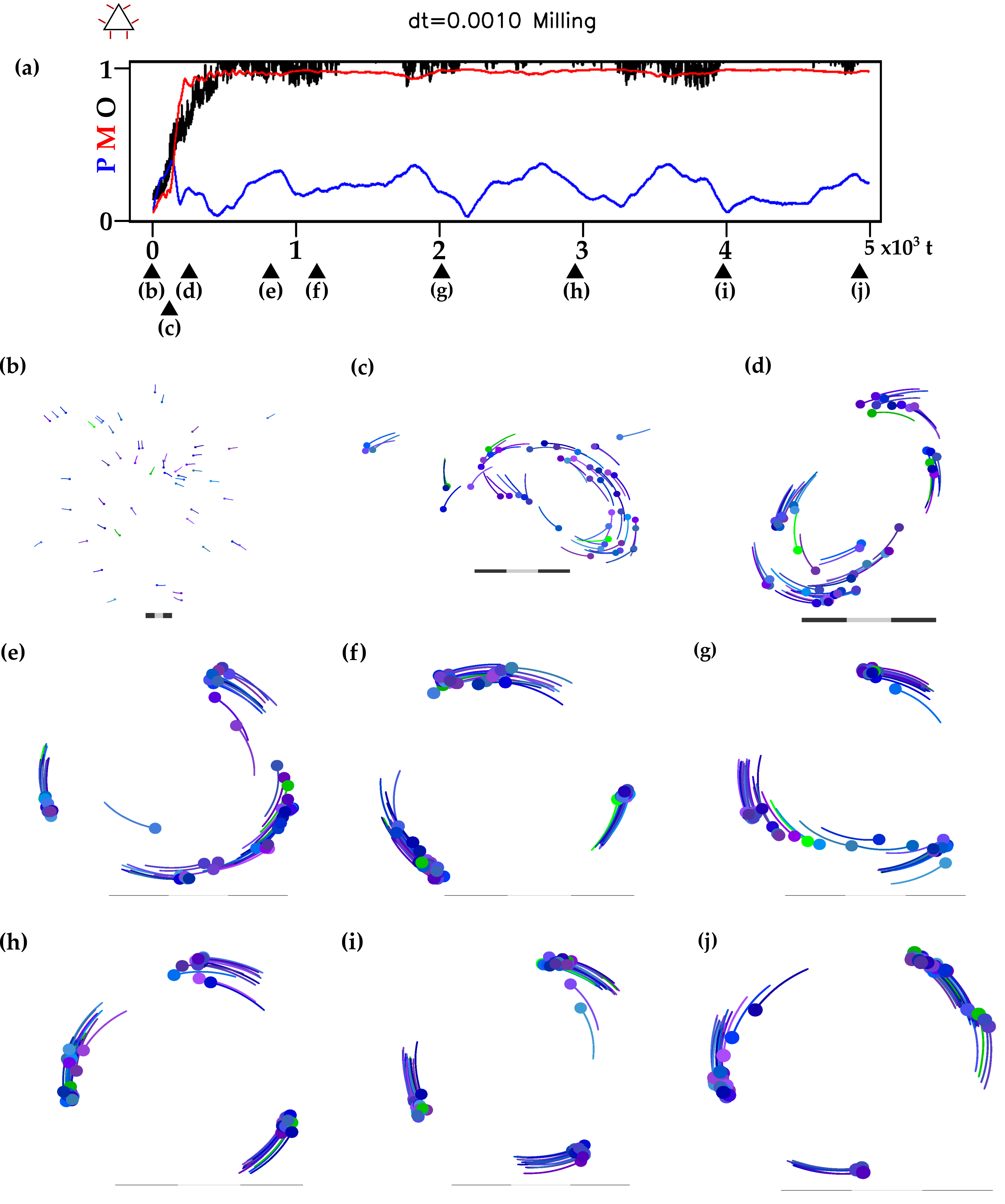}
  \caption{\dtcaption{0.001}{Omniscient}, Video S4 (\url{https://youtu.be/r77__ozLFvc}).}
\end{figure}
\begin{figure}[hb]
  \centering
  \includegraphics[width=1\textwidth]{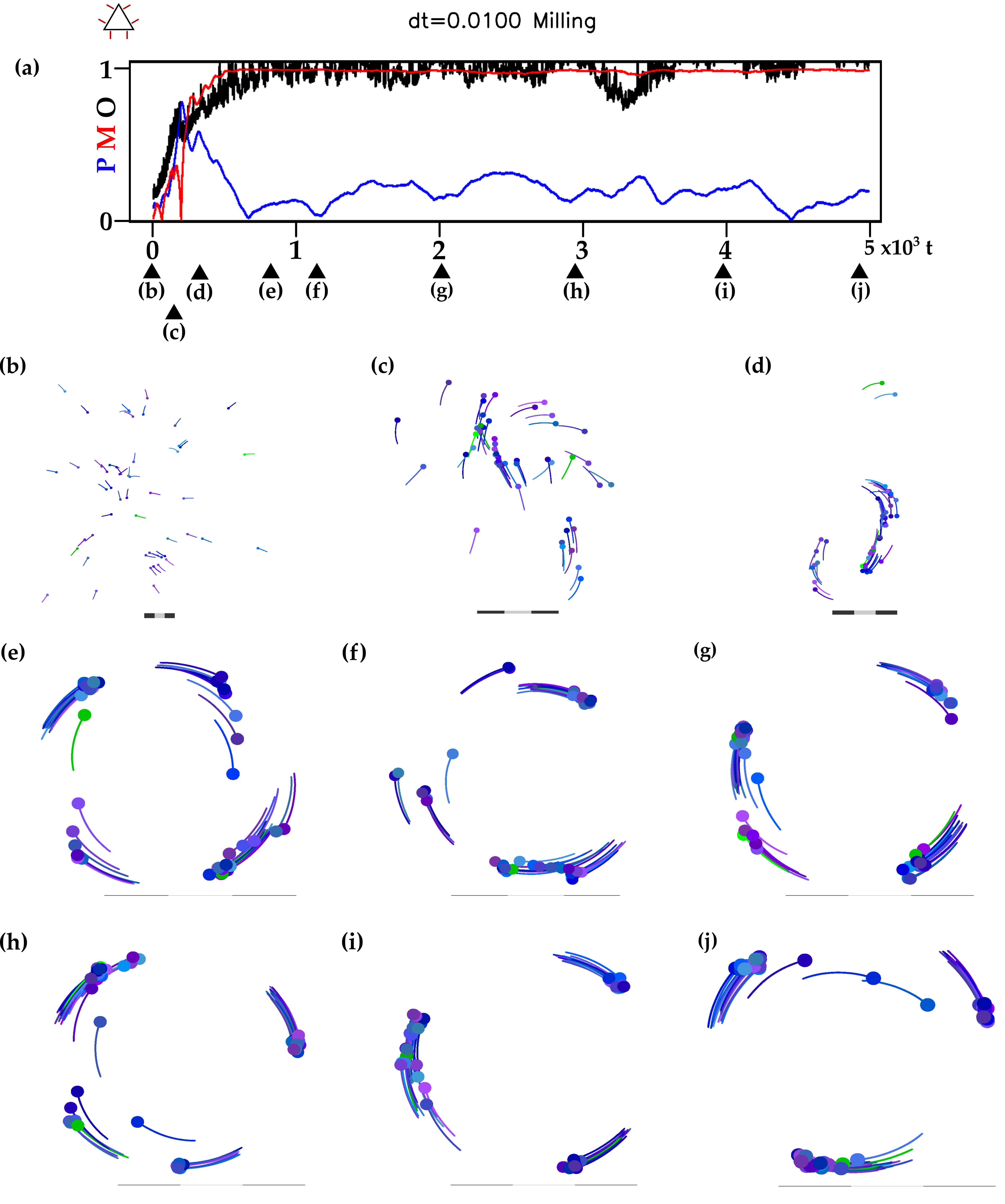}
  \caption{\dtcaption{0.01}{Omniscient}, Video S5 (\url{https://youtube.com/shorts/f_McHAyXwzU}).}
\end{figure}
\begin{figure}[hb]
  \centering
  \includegraphics[width=1\textwidth]{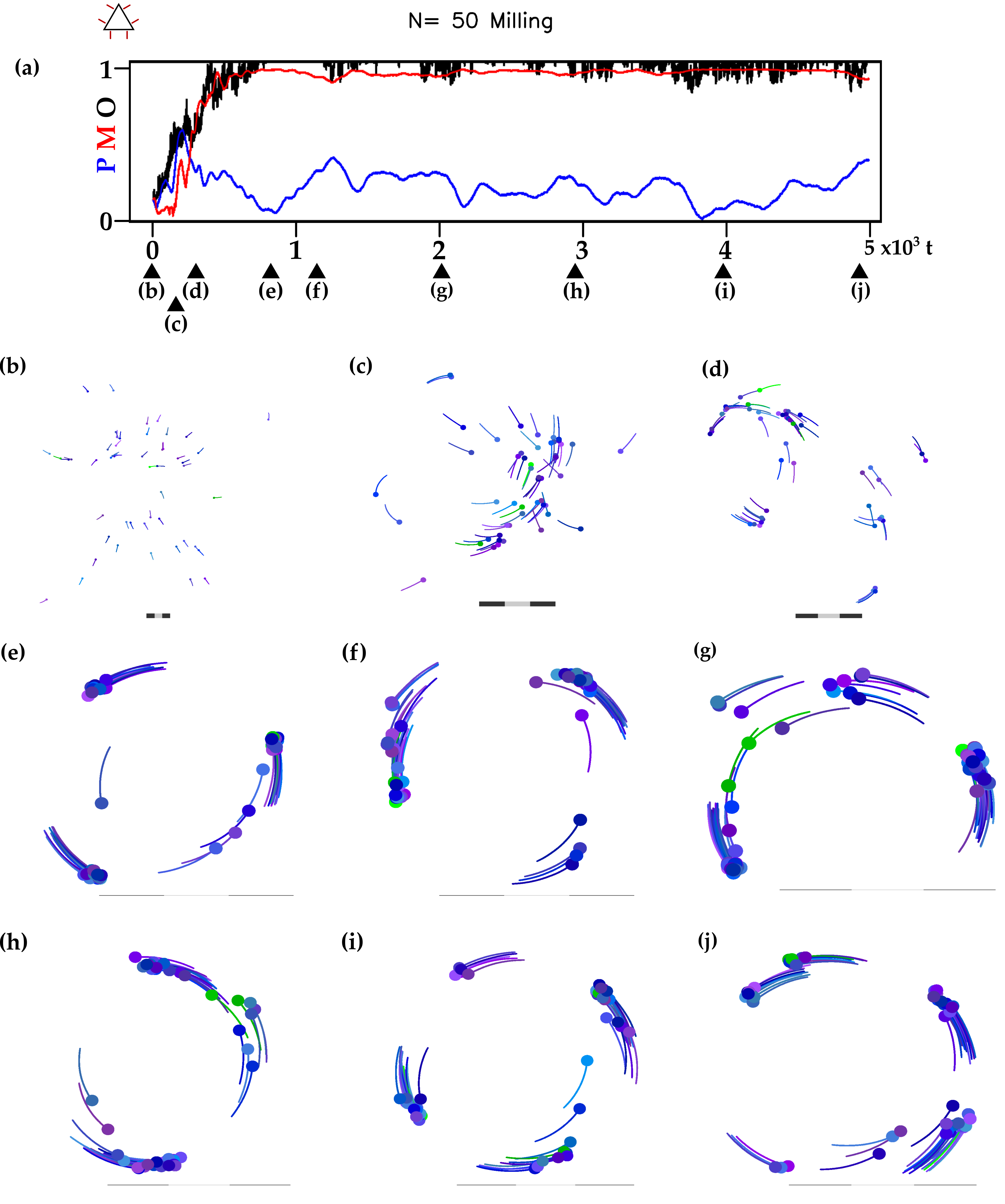}
  \caption{\dtcaption{0.1}{Omniscient}, Video S6 (\url{https://youtu.be/NiVgYHt-Fkc}).}
\end{figure}

\end{widetext}
\clearpage
\begin{widetext}
\sectionmy{Video Snapshots of Phases Reproduction (Fig. 2) ($k_{\eta} = 0.01$, $N=50$)}
\subsectionmy{Visual Model}
\begin{figure}[hb]
  \centering
  \includegraphics[width=0.85\textwidth]{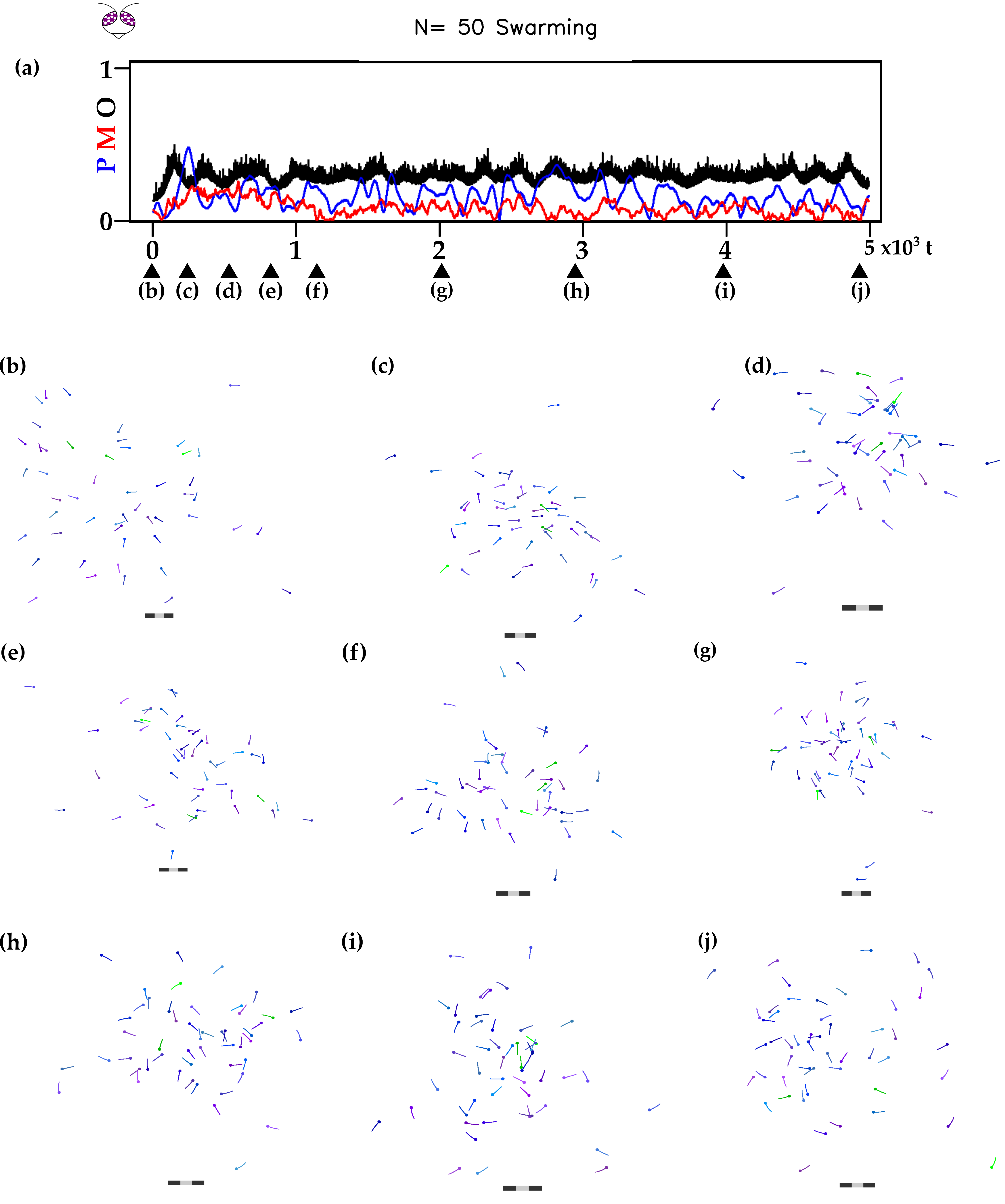}
  \caption{\phasecaption{Swarming}{50}{Visual}{0.06}{0.0}{0.01}{no}, Video 7 (\url{https://youtu.be/U5yeh-7TOUE}).}
\end{figure}
\begin{figure}[hb]
  \centering
  \includegraphics[width=1\textwidth]{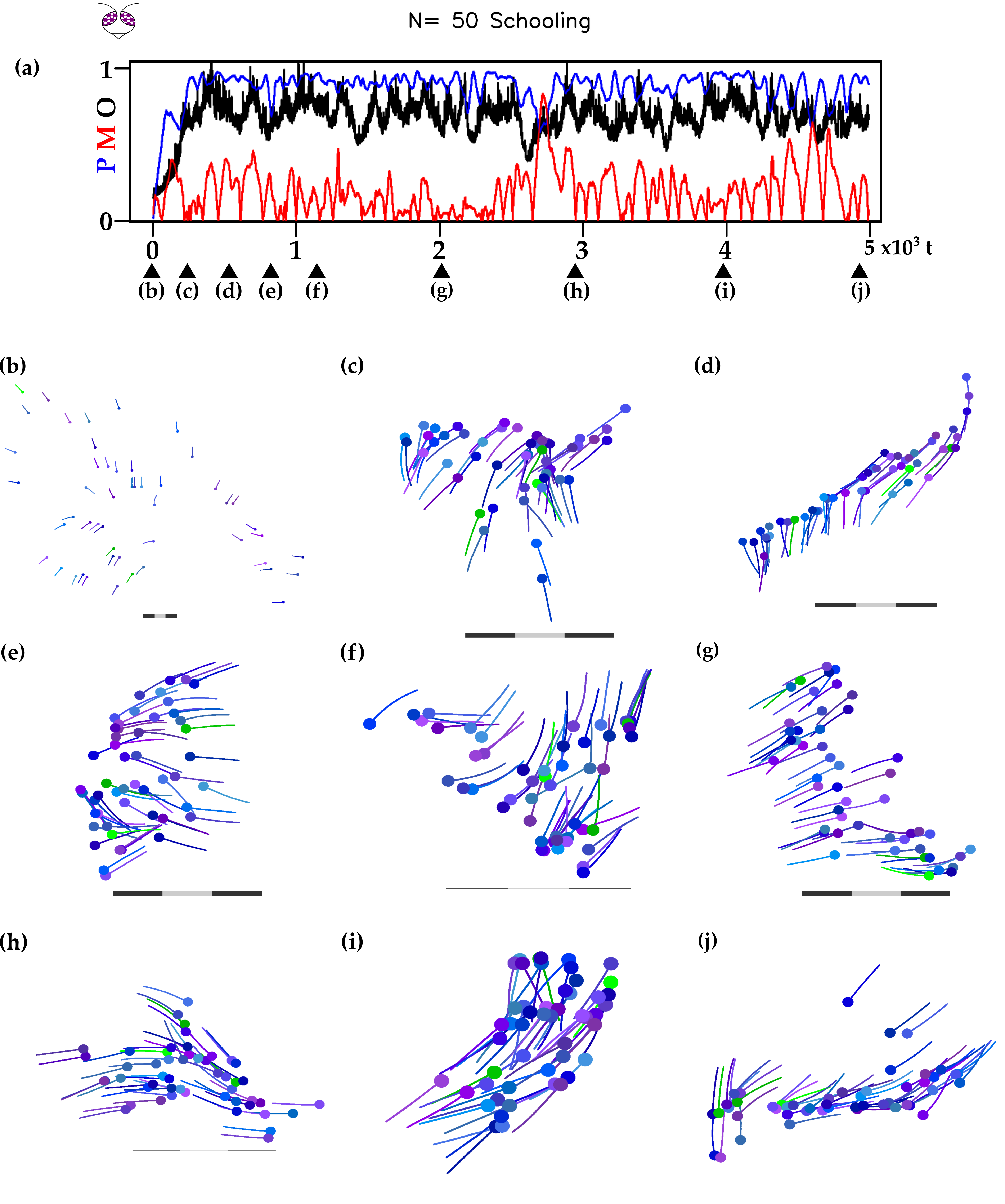}
  \caption{\phasecaption{Schooling}{50}{Visual}{0.06}{0.2}{0.01}{maximum}, Video S8 (\url{https://youtu.be/2sMePYecmzk}).}
\end{figure}
\begin{figure}[hb]
  \centering
  \includegraphics[width=1\textwidth]{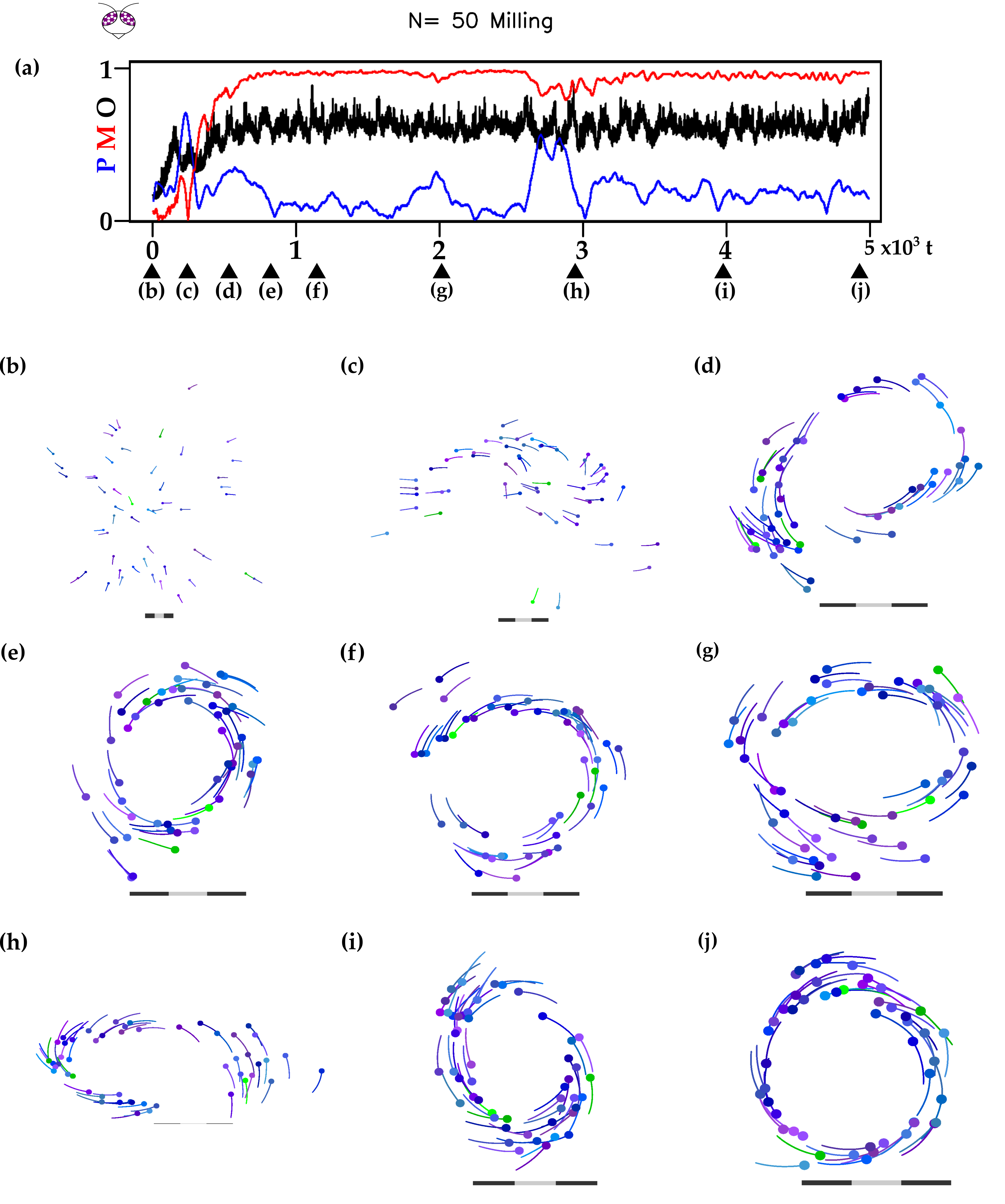}
  \caption{\phasecaption{Milling}{50}{Visual}{0.06}{0.06}{0.01}{maximum}, Video S9 (\url{https://youtu.be/8dFZNKr1Afw}).}
\end{figure}
\end{widetext}
\begin{widetext}
\clearpage
\subsectionmy{Omniscient Model}
\begin{figure}[hb]
  \centering
  \includegraphics[width=0.9\textwidth]{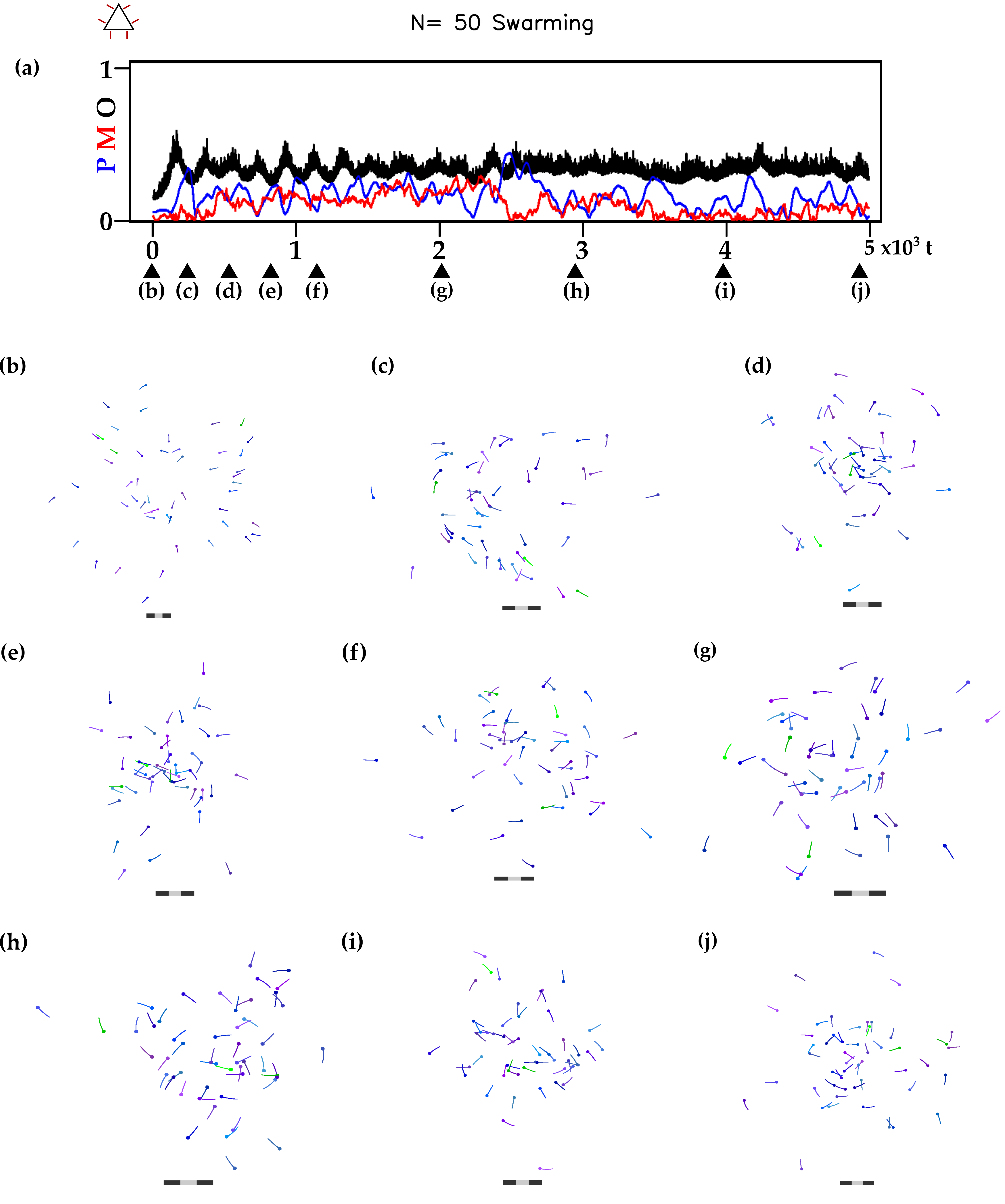}
  \caption{\phasecaption{Swarming}{50}{Omniscient}{0.06}{0.0}{0.01}{no}, Video S10 (\url{https://youtu.be/NgZ8PCJlO2k}).}
\end{figure}
\begin{figure}[hb]
  \centering
  \includegraphics[width=1\textwidth]{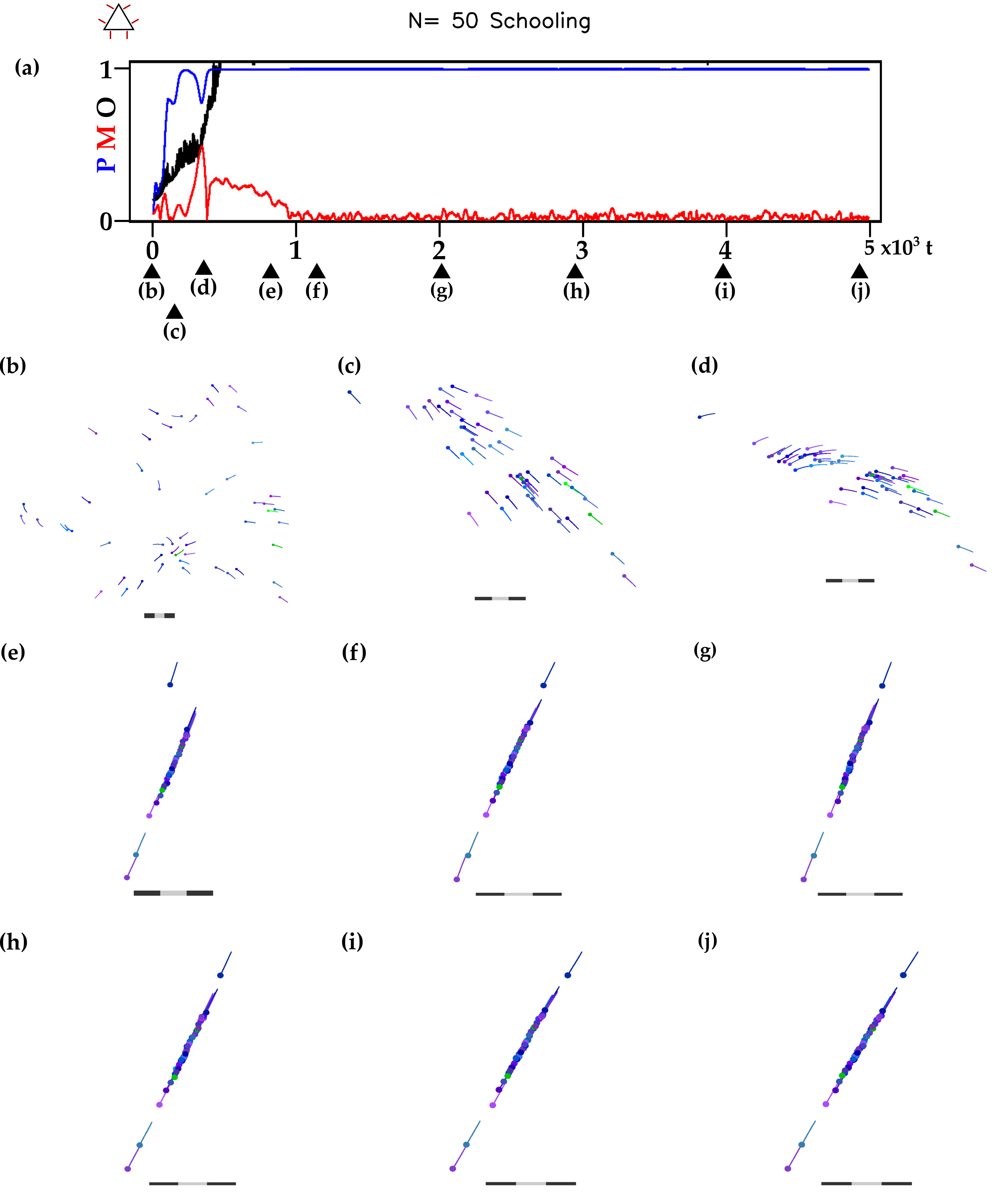}
  \caption{\phasecaption{Schooling}{50}{Omniscient}{0.06}{0.2}{0.01}{maximum}, Video S11 (\url{https://youtu.be/x73EZqu_2a4}).}
\end{figure}
\begin{figure}[hb]
  \centering
  \includegraphics[width=1\textwidth]{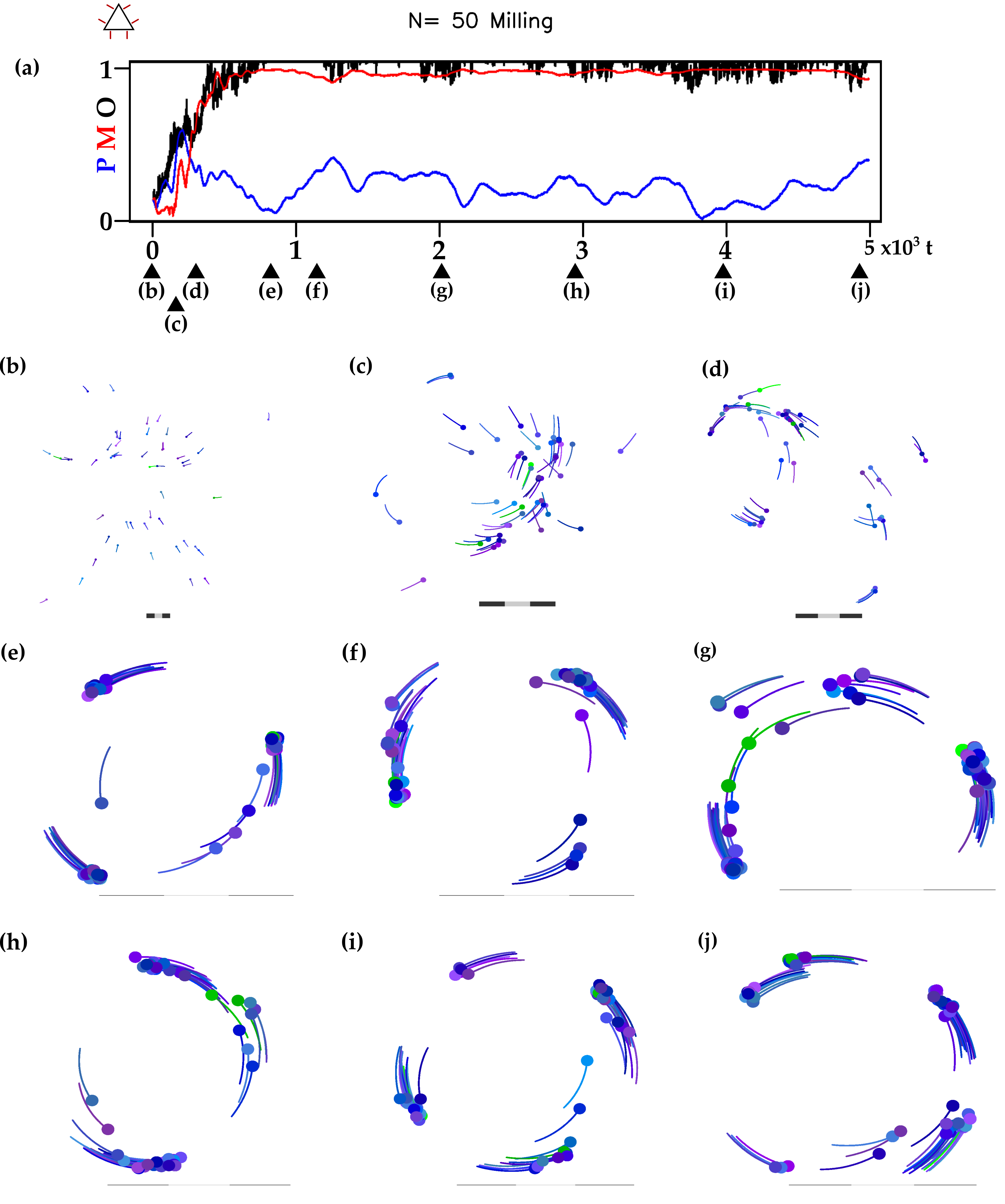}
  \caption{\phasecaption{Milling}{50}{Omniscient}{0.06}{0.06}{0.01}{maximum}, Video S12 (\url{https://youtu.be/CWpTSvr_zBk}).}
\end{figure}
\clearpage
\sectionmy{Video Snapshots of Bistable Phase Visual model (Fig. 4)}
\begin{figure}[hb]
  \centering
  \includegraphics[width=0.9\textwidth]{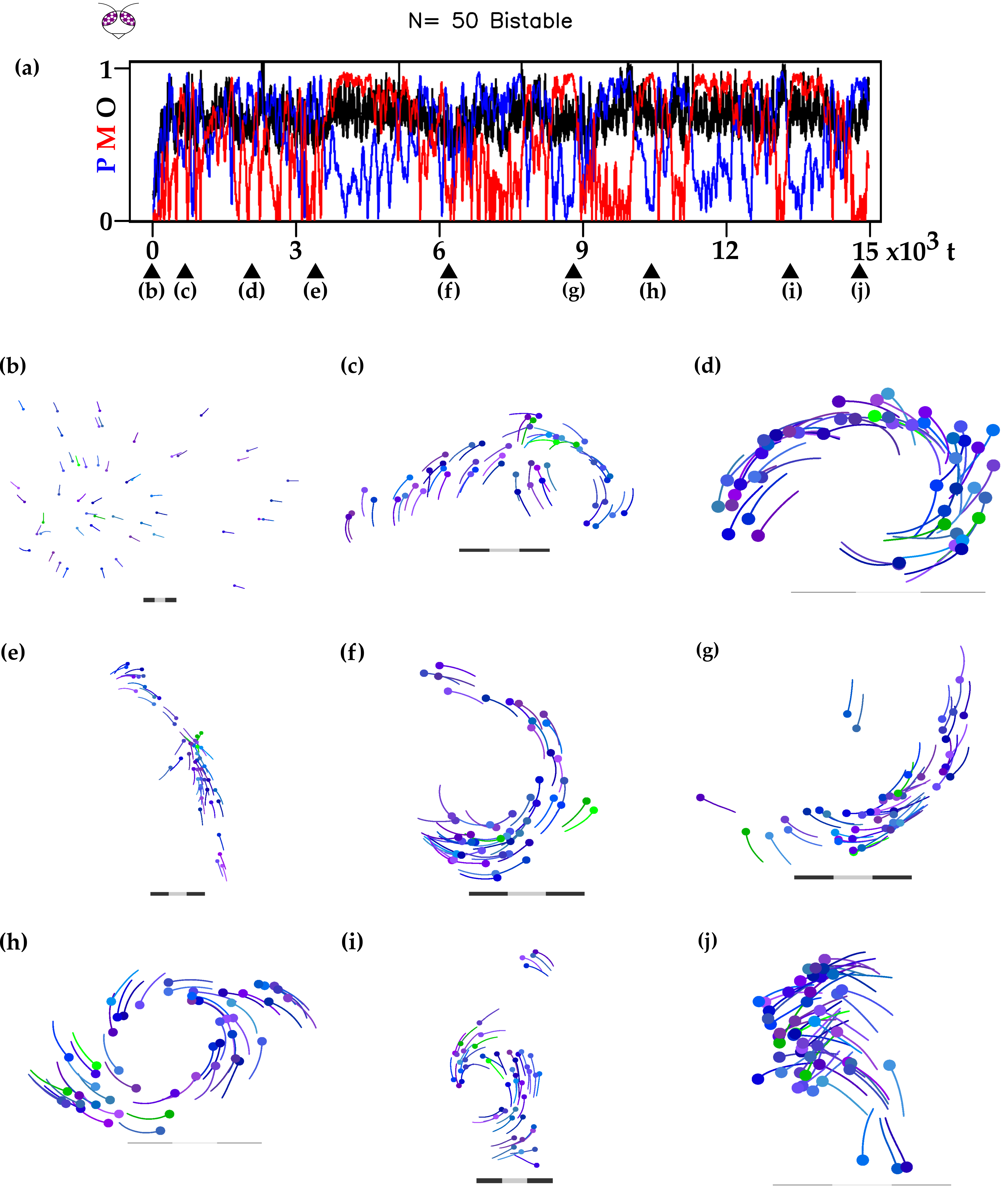}
  \caption{\phasecaption{Bistable}{50}{Visual}{0.1}{0.2}{0.01}{maximum}, Video S13 (\url{https://youtu.be/JVCp65SVX2Q}).}
\end{figure}

\end{widetext}
\clearpage
\renewcommand{\refname}{Supplemental References}

\end{document}